\documentclass[a4paper,fleqn,usenatbib,useAMS]{mnras}
\usepackage{newtxtext,newtxmath}
\usepackage[T1]{fontenc}
\usepackage{ae,aecompl}
\usepackage{graphicx}
\usepackage{natbib}
\usepackage{amssymb}
\usepackage{amsmath}
\usepackage{xcolor}

\newcommand{\msun}{\mbox{${\rm M}_{\odot}$}}

\def\lesssim{\lower.5ex\hbox{$\; \buildrel < \over \sim \;$}}
\def\gtrsim{\lower.5ex\hbox{$\; \buildrel > \over \sim \;$}}

\title[The Nature of Massive Transition Galaxies]{The Nature of Massive Transition Galaxies in CANDELS, GAMA, and Cosmological Simulations}

\author[V. Pandya et al.] {Viraj Pandya$^{1,2,3}$,
Ryan Brennan$^2$, 
Rachel S. Somerville$^2$, 
Ena Choi$^2$,
\newauthor Guillermo Barro$^4$, 
Stijn Wuyts$^5$, 
Edward N. Taylor$^6$,
Peter Behroozi$^{4,7}$,
\newauthor Allison Kirkpatrick$^8$,
Sandra M. Faber$^3$, 
Joel Primack$^{9}$,
David C. Koo$^3$,
\newauthor Daniel H. McIntosh$^{10}$,
Dale Kocevski$^{11}$, 
Eric F. Bell$^{12}$, 
Avishai Dekel$^{13}$,
\newauthor Jerome J. Fang$^{3,14}$,
Henry C. Ferguson$^{15}$,
Norman Grogin$^{15}$,
Anton M. Koekemoer$^{15}$,
\newauthor Yu Lu$^{16}$,
Kameswara Mantha$^{10}$,
Bahram Mobasher$^{17}$,
Jeffrey Newman$^{18}$,
\newauthor Camilla Pacifici$^{19}$,
Casey Papovich$^{20}$,
Arjen van der Wel$^{21}$,
Hassen M. Yesuf$^3$\\
\\Affiliations are listed at the end of this paper.
}

\date{Accepted ???. Received ??? in original form ???}

\pubyear{2016}

\begin{document}
\label{firstpage}
\pagerange{\pageref{firstpage}--\pageref{lastpage}}
\maketitle

\begin{abstract}
We explore observational and theoretical constraints on how galaxies might transition between the ``star-forming main sequence" (SFMS) and varying ``degrees of quiescence" out to $z=3$. Our analysis is focused on galaxies with stellar mass $M_*>10^{10}M_{\odot}$, and is enabled by GAMA and CANDELS observations, a semi-analytic model (SAM) of galaxy formation, and a cosmological hydrodynamical ``zoom in" simulation with momentum-driven AGN feedback. In both the observations and the SAM, transition galaxies tend to have intermediate S{\'e}rsic indices, half-light radii, and surface stellar mass densities compared to star-forming and quiescent galaxies out to $z=3$. We place an observational upper limit on the average population transition timescale as a function of redshift, finding that the average high-redshift galaxy is on a ``fast track" for quenching whereas the average low-redshift galaxy is on a ``slow track" for quenching. We qualitatively identify four physical origin scenarios for transition galaxies in the SAM: oscillations on the SFMS, slow quenching, fast quenching, and rejuvenation. Quenching timescales in both the SAM and the hydrodynamical simulation are not fast enough to reproduce the quiescent population that we observe at $z\sim3$. In the SAM, we do not find a clear-cut morphological dependence of quenching timescales, but we do predict that the mean stellar ages, cold gas fractions, SMBH masses, and halo masses of transition galaxies tend to be intermediate relative to those of star-forming and quiescent galaxies at $z<3$.
\end{abstract}


\section{Introduction}\label{sec:intro}
It is well known that there exists a bimodality in galaxy colors and star formation rates (SFRs) out to at least $z\approx3$ \citep[e.g.,][]{strateva01,brinchmann04,bell04,baldry04,faber07,wyder07,blanton09,whitaker11,straatman16}. In particular, the distribution of galaxy SFRs and colors splits into: (1) a ``red sequence'' of quiescent galaxies that host very little, if any, ongoing star formation and that are dominated by a relatively old stellar population, and (2) a ``blue cloud'' of star-forming galaxies that are actively forming new stars and are dominated by a young stellar population. Evidence suggests that the typical SFRs, colors, and other properties of these two populations change significantly as a function of cosmic time, implying evolution both within and between the two populations \citep[e.g., see][and references therein]{madau14}. Furthermore, the fraction of all galaxies that are quiescent has increased with cosmic time, and this increase in the quiescent fraction happens earlier for more massive galaxies \citep[this is known as ``cosmic downsizing"; e.g.,][and references therein]{cowie96,noeske07b,fontanot09}.

It was thought since at least the 1970s that there may exist a third population of ``transient'' galaxies that are transitioning between what we now call the blue cloud and the red sequence, although such work was mostly restricted to dense environments such as clusters \citep[e.g.,][]{vandenbergh76,butcheroemler78}. A well-known example of such galaxies are the classical post-starburst or ``K+A'' (or more restrictively, ``E+A") galaxies, which are predominantly old stellar systems that contain some A-type stars due to a recently truncated starburst \citep[e.g.,][]{dressler83}. However, such post-starburst galaxies are now confirmed to be rare, at least in the local Universe \citep[e.g.,][]{quintero04,wild09,yesuf14,pattarakijwanich14,mcintosh14}. Despite the higher observed number densities of post-starburst galaxies at $z>1$ \citep[e.g.,][]{whitaker12}, it is still not at all clear that this population can by itself account for the dramatic build-up of the red sequence across cosmic time \citep[but see][for an alternative view]{wild16}. Hints of a broad and general framework for the large-scale transition of galaxies between different populations did not clearly and explicitly begin to emerge until the early 2000s when statistical samples of galaxies became available through the advent of large astronomical surveys \citep[e.g.,][]{colless01,strateva01,strauss02}.

\citet{bell04} explicitly studied what they called the ``gap'' population (between the red sequence and blue cloud, defined using the classical color-magnitude diagram) and found that by turning off star formation in some small fraction of blue galaxies, such galaxies would fade across the gap, join the red sequence, and at least qualitatively explain the build-up of the red population since $z\sim1$. Around the same time, \citet{baldry04} quantitatively studied the bimodal color-magnitude distribution of galaxies and found that, in the local Universe, the red and blue populations could be adequately modeled as the sum of two Gaussians, implying no need for such a ``gap'' population and therefore suggesting that all galaxies transition on extremely fast timescales. \citet{faber07} quantitatively demonstrated the build-up of the red sequence since $z\sim0.7$ and qualitatively explored the different evolutionary pathways that
galaxies could follow in order to become truly red-and-dead galaxies.

The idea of the classical ``green valley'' population was born and systematically explored in the seminal 2007 series of papers celebrating the advent of the ultraviolet \textit{Galaxy Evolution Explorer (GALEX)} space-based telescope \citep{wyder07,martin07,schiminovich07,salim07}. \citet{wyder07} showed that the ``gap'' population studied by \citet{bell04} became much more pronounced in the NUV-optical color-magnitude diagram (i.e., using $NUV-r$ color rather than $u-r$ or $g-r$ color) because the blackbody emission from young stars peaks in the NUV, allowing for much finer constraints on the recent star formation histories (SFHs) of galaxies. \citet{martin07} and \citet{goncalves12} found that the recent SFHs of green valley galaxies, and their quenching timescales in particular, were indeed consistent with the build-up of the red sequence implied by the cosmic evolution of the blue and red galaxy luminosity functions.

The classical green valley population clearly has major implications for theories of galaxy evolution, but in the decade since its discovery, many possible caveats and uncertainties have been raised that have led to inconsistencies and confusion in the existing literature \citep[see also the extensive discussions in][]{salim14,schawinski14}. There are concerns that the classical green valley population mostly comprises dusty star-forming galaxies \citep[e.g.,][]{brammer09,cardamone10}, or blue and red galaxies that have been scattered into this intermediate color range due to measurement errors \citep[the ``purple valley" interpretation; see][]{mendez11}. Furthermore, since many low-redshift studies have found that classical green valley galaxies tend to be composite bulge plus disk systems, it is said that their intermediate colors are merely the result of superimposing a red bulge onto a blue disk \citep[e.g.,][]{dressler15}. While intriguing, this latter interpretation does not adequately explain how the ``superimposed" red bulges grew in the first place, why they are preferentially hosted by galaxies in the green valley and the red sequence, and what the physical relationship is between bulge formation histories and star formation histories. 

Two ways to help address these concerns and refine our understanding of the evolutionary role of the green valley population are to: (1) extend our study out to high redshift, where the red sequence is not yet built up and the vast majority of galaxies are forming stars at higher rates than locally, and (2) use the more physically-motivated SFR--stellar mass diagram to identify stragglers below the tight ``star-forming main sequence" of blue galaxies \citep[SFMS; e.g.,][]{noeske07}. Previous studies of the green valley population were limited to low- and intermediate-redshift because it is only for these relatively nearby galaxies that there exists an abundance of spectroscopic and imaging data with relatively high physical resolution \citep[e.g.,][]{martin07,salim10,mendez11,fang12,salim12,goncalves12,pan13,pan14,schawinski14,mcintosh14,smethurst15,haines15}.

In this paper, we will extend the study of the green valley population (what we call the transition population) out to $z=3$ based on the wealth of new high spatial resolution observations taken with the \textit{Hubble Space Telescope} Wide Field Camera 3 (\textit{HST}/WFC3) as part of the Cosmic Assembly Near-infrared Deep Extragalactic Legacy Survey \citep[CANDELS;][]{candels1,candels2}. We will also self-consistently analyze predictions from a semi-analytic model of galaxy formation in the same way as the observations, and compare the observational and semi-analytic results to those that we obtain from a state-of-the-art hydrodynamical simulation. These comparisons can help constrain the implementation of physical processes in models and motivate future studies of the transition galaxy population in a cosmological context. Specifically, the questions that we will aim to address in this paper are: 
\begin{enumerate}
\item How do the structure and morphology of transition galaxies compare to those of star-forming and quiescent galaxies as a function of redshift?
\item How do the relative fractions of galaxies that are star-forming, transitioning, and quiescent evolve with redshift, and what are the implications for the average population transition timescale as a function of redshift?
\item Where do the models agree and disagree with the observations, and how might the models be improved?
\item What is the physical origin of transition galaxies in the models and do their quenching timescales have a clear-cut morphological dependence?
\item What other physical properties are predicted by the models to be useful for robustly identifying transition galaxies at a range of redshifts with future observations?
\end{enumerate}

This paper is organized as follows. In \autoref{sec:datadesc}, we describe the observations and in \autoref{sec:samdesc} we describe the semi-analytic model. In \autoref{sec:methods} we explain our methods, and in \autoref{sec:results} we present our results. After an observational discussion in \autoref{sec:discobs} and a theoretical discussion in \autoref{sec:disctheory}, we summarize in \autoref{sec:summary}. Throughout the paper, we assume $H_0 = 67.8$ km s$^{-1}$ Mpc$^{-1}$, $\Omega_m=0.307$, and $\Omega_{\Lambda}=0.693$ following \citet{planck14}.

\section{Observations}
\label{sec:datadesc}
\subsection{CANDELS}
The backbone of our study is $HST$/WFC3 imaging taken as part of CANDELS \citep{candels1,candels2}. The CANDELS data span five different fields which collectively add up to $\sim0.22$ deg$^2$. This large area helps to minimize the effects of cosmic variance \citep[e.g., see][]{somerville04}. The five CANDELS fields and their associated data description papers are: COSMOS (Nayyeri et al., in preparation), EGS (Stefanon et al., in preparation), GOODS-N (Barro et al., in preparation), GOODS-S \citep{guo}, and UDS \citep{galametz}. 

A major advantage of CANDELS is that the galaxies are selected in the near-IR F160W ($H$) band. This allows us to probe the rest-frame UV-optical spatial and SED features of galaxies out to $z\sim3$ with unprecedentedly high resolution. In what follows, we will briefly describe the derivation of the most relevant physical parameters in the CANDELS catalogs.\footnote{All CANDELS catalogs are available at the Rainbow Database: \url{http://arcoiris.ucolick.org/Rainbow_navigator_public/}} For in-depth details about the data processing and source catalog creation for each CANDELS field, we refer the reader to the five data description papers cited above. Our overview below applies uniformly to all five CANDELS fields.

First, the template-fitting method \citep[TFIT;][]{tfit1,tfit2} was used to merge multi-wavelength (UV to near-IR) observations with significantly different spatial resolutions, and construct the observed-frame multi-wavelength photometric catalog. Photometric redshifts were derived using the Bayesian framework described in \citet{dahlen}; this method combines the posterior redshift probability distributions from several independent codes to improve precision and reduce outliers. Spectroscopic redshifts were used where available and reliable; $HST$/WFC3 grism redshifts from \citet{morris15} were used for GOODS-S.

Rest-frame UV-optical-NIR photometry was derived by fitting the observed-frame SED with a set of templates using the EAZY code \citep[][and Kocevski et al., in preparation]{brammer08}. The method for computing stellar masses is described in \citet{santini15}, and a critical assessment of the method, including the possible contribution of nebular emission to stellar mass estimates, is given in \citet{mobasher15}. Several independent codes \citep[including FAST;][]{kriek09fast} were used to derive stellar masses under a set of fixed assumptions, but with room for some variation such as assumed SFH parameterizations. Although the underlying data are the same, the use of several different SED codes and assumptions allows one to test the impact of systematic errors and to analyze the precision of estimated stellar masses. For our study, we use physical properties that were derived assuming the following: \citet{bc03} stellar population synthesis models, \citet{chabrier} initial mass function (IMF), exponentially-declining SFHs, solar metallicity, and the \citet{calzetti00} dust attenuation law.

For galaxies that are detected at 24$\mu$m with \textit{Spitzer}-MIPS, the total IR luminosity ($L_{\rm TIR}$) was computed using the mapping from 24$\mu$m flux to $L_{\rm TIR}$ given in \citet{elbaz11}. In some cases, galaxies detected at $24\mu$m also have significantly detected (and deblended) counterparts in far-IR \textit{Herschel}-SPIRE imaging at 250, 350 and 500 $\mu$m; for these, we instead use their best-fitting IR templates to determine $L_{\rm TIR}$ \citep[][]{perezgonzalez10,barro11}. Both of these techniques (24$\mu$m-based mappings and IR template fitting) have two built-in assumptions: (1) there is minimal, if any, redshift evolution of the intrinsic IR SEDs of galaxies across a rather large redshift range (limited to $0<z<3$ for our study), and (2) emission from an obscured active galactic nucleus (AGN) does not contribute significantly to the 24$\mu$m flux \citep[these topics are discussed extensively in][]{elbaz11,wuyts11,barro11}. We will return to the impact of dust-obscured AGN near the end of this subsection.

SFRs were derived for galaxies according to a ladder of SFR indicators based on the prescriptions given in \citet{wuyts11} and \citet{barro11}. By default, every galaxy has an estimate of SFR$_{\rm UV}$ derived from its SED-based rest-frame NUV luminosity at 2800\AA, $L_{2800}$. We use the NUV rather than the FUV (1500\AA) because the blackbody emission of young stars peaks in the NUV. We correct this UV-based SFR for dust attenuation by assuming the \citet{calzetti00} dust attenuation curve:
\begin{equation}
\rm SFR_{\rm UV,corr}\;[M_{\odot} \;\rm yr^{-1}] = \rm SFR_{\rm UV}\times10^{0.4\times1.8\times A_V} \quad,
\end{equation}
where SFR$_{\rm UV} = 3.6\times10^{-10}\times L_{2800}/L_{\odot}$ assuming a \citet{chabrier} IMF as in \citet{wuyts11}. In the exponent, $A_V$ is the SED-based optical attenuation output by FAST \citep{kriek09fast}, and the factor of 1.8 corresponds to the \citet{calzetti00} attenuation curve value at 2800\AA.

For galaxies that are also detected in mid-IR (and possibly far-IR) imaging and thus have $L_{\rm TIR}$ measurements as described above, we can alternatively compute the total SFR
as the sum of the unobscured, non-dust-corrected SFR$_{\rm UV}$ and the obscured SFR$_{\rm IR}$ \citep[as described in][]{wuyts11,barro11}:
\begin{equation}
\rm SFR_{\rm UV+IR}\;[M_{\odot} \;\rm yr^{-1}]  = \rm SFR_{\rm UV} + 1.09\times10^{-10}\times L_{\rm TIR}/L_{\odot}\quad.
\end{equation}
We adopt SFR$_{\rm UV+IR}$ as our standard indicator for all galaxies that have $L_{\rm TIR}$ measurements, and SFR$_{\rm UV,corr}$ otherwise. It is very interesting to consider the impact of excluding mid- and far-IR contributions by instead using dust-corrected SFR$_{\rm UV,corr}$ only, even if SFR$_{\rm UV+IR}$ is available. If we re-run our entire analysis in this paper using only SFR$_{\rm UV,corr}$, then our results are slightly perturbed but the main conclusions do not change. Similarly, we also re-ran our entire analysis using the UV-optical-NIR SED-based SFR$_{\rm SED}$ output by FAST \citep{kriek09fast} for every galaxy; these fits do not use bandpasses beyond the $8\mu$m channel of \textit{Spitzer}-IRAC. Again, our exact quantitative results are perturbed but our conclusions do not change. Interestingly, the SFMS in the sSFR-M$_*$ diagram is more negatively sloped when using SFR$_{\rm SED}$ compared to both SFR$_{\rm UV,corr}$ and SFR$_{\rm UV+IR}$. However, when allowing this slope to be a free parameter, our results are insensitive to the choice of SFR indicator.

One potential issue with UV-based SFRs can arise when a galaxy is not detected in the observed frame filter corresponding to rest-frame 2800\AA\ at its redshift (or in either of the two adjacent filters). EAZY \citep{brammer08} may then extrapolate its NUV luminosity based on detections at significantly different wavelengths, leading to highly uncertain $L_{2800}$ and thus unreliable SFR$_{\rm UV,corr}$ values. If the SFR$_{\rm UV,corr}$ values of intrinsically star-forming or transition galaxies are underestimated, then the fraction and number density of quiescent galaxies at high redshift might be artificially boosted. The inverse is not as much of a problem because upper limits on $L_{2800}$ non-detections naturally set a floor on SFR$_{\rm UV,corr}$, below which we cannot detect quiescent galaxies anyway. We verified that in each redshift slice for a given CANDELS field, the majority of galaxies (usually $>90\%$, at worst $\sim70\%$) that make it past our selection cuts are indeed detected in the rest-frame filter corresponding to 2800\AA\ at their respective redshifts. For the minority of galaxies that are not detected at 2800\AA, their SFR$_{\rm UV,corr}$ is rarely lower than the SFR$_{\rm UV,corr}$ of the average robustly NUV-detected galaxy. This means that it is unlikely that our quiescent fractions at high redshift are significantly overestimated due to rest-frame NUV non-detections. 

On a related note, the presence of an obscured AGN may boost an otherwise quiescent galaxy's $24\mu$m-based SFR$_{\rm IR}$ and cause that galaxy to instead become classified as a star-forming or transition galaxy. This can make it difficult to test whether the transition region might indeed be an evolutionary bridge between the SFMS and the quiescent region. We have used the procedure described in \citet[][submitted]{kirkpatrick13,kirkpatrick15} to assign a mid-IR luminosity AGN contribution fraction to each CANDELS galaxy (based on the $8\mu$m/$3.6\mu$m versus $250\mu$m/$24\mu$m diagnostic diagram). We find that it is unlikely that obscured AGN are preferentially boosting the SFRs of transition and quiescent galaxies, and thus that contamination by obscured AGN does not significantly affect our results. A full analysis of the demographics of this obscured AGN population in the context of the transition region will be presented in forthcoming work.

Lastly, structural parameters were derived for every galaxy using GALFIT \citep{peng}. The fits were done to the \textit{HST}/WFC3 F160W ($H$-band) images \citep{vanderwel12} using a global S{\'e}rsic model. We emphasize that there is a difference between fitting structural properties to \textit{H}-band light images instead of stellar mass images. In this work, we use half-light radii (i.e., the semi-major axis radius), as opposed to half-mass radii. Similarly, our S{\'e}rsic indices give information about the $H$-band light distribution rather than the stellar mass distribution for each galaxy. Although studies suggest that adopting mass-based rather than light-based structural parameters would not significantly change our results \citep[e.g.,][]{szomoru13,lang14}, in the future it will be important to revisit this claim. The original GALFIT measurements for S{\'e}rsic index were allowed to run from $n=0$ to $n=8$, where $n=8$ corresponds to very compact light profiles; for ease of interpretation, we set all $n>4$ to the classic de Vaucouleurs index, $n=4$.

Since our CANDELS observations span a large range in redshift and we wish to compare our results to those that we obtain from the low-redshift GAMA survey, it is important to apply morphological k-corrections to the structural measurements of CANDELS galaxies. \citet{vanderwel14} provide a formula for converting half-light radii from observed-frame $H$-band to rest-frame 5000\AA. We find that our results are not significantly affected by the conversion of \citet{vanderwel14}. Therefore, in this paper, we will show our results in terms of observed-frame $H$-band S{\'e}rsic indices and half-light radii. In the future, it will be important to revisit this non-trivial task of morphological k-corrections.

In order to ensure high sample completeness and robust structural measurements, we make the following selection cuts: F160W apparent magnitude $< 25$, stellar mass $M_*>10^{10}M_{\odot}$, and F160W GALFIT quality flag = 0 (good fits only). The total fraction of galaxies cut out by requiring good GALFIT structural measurements is $<15\%$; we found that these discarded galaxies do not occupy a special limited subspace of the sSFR-M$_*$ or $UVJ$ diagrams. We find that we are complete to star-forming galaxies down to at least $M_*=10^9M_{\odot}$ out to $z=3$ (the redshift limit for this paper), and that our main analysis sample is also complete down to $M_*\approx10^{10}M_{\odot}$ out to $z=3$ \citep[see also][for a discussion of completeness limits in CANDELS]{newman12}.

\subsection{GAMA}
The volume of the CANDELS fields is extremely small below $z\sim0.5$, so we cannot extend our CANDELS analysis much below this redshift. However, there are $\sim3.5$ Gyr between our CANDELS low redshift limit, $z=0.5$, and the local Universe at $z\sim0.1$. In order to connect our results from CANDELS at $0.5<z<3.0$ with the nearby Universe at $z\sim0.1$, we augment our analysis with Data Release 2 (DR2) from the Galaxy And Mass Assembly survey \citep[GAMA;][]{liske15}. GAMA is a large (144 sq. deg.) survey that builds on the legacy of the Sloan Digital Sky Survey \citep[SDSS;][]{york00} and the Two-degree Field Galaxy Redshift Survey \citep[2dF-GRS;][]{colless01}. The ``main galaxy survey" component of GAMA goes two magnitudes deeper ($r<19.8$ mag) than that of SDSS \citep{strauss02} while maintaining very high ($\gtrsim98$\%) spectroscopic completeness. Like CANDELS, GAMA has also inherited a rich supplementary multi-wavelength dataset running from 1 nm to 1 m \citep{liske15}. The backbone of GAMA is deep optical spectroscopy with the Anglo-Australian Telescope (AAT), and its multi-wavelength catalogs are bolstered by collaborations with several other independent surveys \citep[for a review, see][]{driver11}. 

Specifically, the GAMA DR2 public catalog contains 72,225 objects in three unique GAMA fields: two 48 sq. deg. fields with $r<19.0$ mag limits, and one 48 sq. deg. field with an $r<19.4$ mag limit. This gives a total survey volume of 144 sq. deg; see \citet{baldry10} for more information about survey target selection.

Here we give only a brief overview of the relevant physical properties available in the GAMA DR2 public catalog. We adopt bulk flow-corrected redshifts \citep{baldry12}. The rest-frame photometry and stellar masses were derived from SED fitting as described in \citet{taylor11}, though here we make use of the \textit{VST} VIKING near-IR data discussed in \citet{taylor15} as well. We applied aperture corrections to the stellar masses and rest-frame photometry, as suggested in \citet{taylor11}, to account for the fraction of flux that falls outside the $r$-band aperture used for aperture-matched photometry in Source Extractor \citep{bertin96}. 

Unlike with CANDELS, the high spectroscopic completeness of GAMA affords us H$\alpha$-based star formation rates, SFR$_{\rm H\alpha}$. The SFR$_{\rm H\alpha}$ are based on extinction-corrected H$\alpha$ line luminosities \citep{gunawardhana13}. We converted the original SFR$_{\rm H\alpha}$ from a Salpeter IMF normalization to a Chabrier IMF normalization to be consistent with CANDELS. The SFR$_{\rm H\alpha}$ measurements probe SFRs on timescales of $\sim10$ Myr, unlike the $\sim100$ Myr timescales probed by our CANDELS UV+IR based SFRs. While we are thus more sensitive to low level recent star formation in GAMA with SFR$_{\rm H\alpha}$, this also has the ``drawback" of being more sensitive to stochastic variations in the SFR on shorter timescales \citep[see the relevant GAMA paper by][which focuses only on typical star-forming galaxies]{davies16}. Another caveat is that the nuclear region of a galaxy, which is the region that the GAMA spectral fiber and thus H$\alpha$ line luminosity probes, is not necessarily representative of the galaxy as a whole; there is roughly $\sim0.15$ dex of additional scatter expected due to the conversion from fiber SFR$_{\rm H\alpha}$ to global SFR$_{\rm H\alpha}$ \citep{richards16}.

Structural properties of GAMA galaxies are provided via multi-band measurements using GALFIT \citep{peng}, as described in \citet{kelvin12}. We adopt the GAMA structural fits in the $r$-band; this has the advantage that, like CANDELS, we will be analyzing the structural properties of GAMA galaxies in the same band in which those galaxies were selected (namely, the $r$ band). More importantly, since most of our \textit{H}-band-selected CANDELS galaxies are at $z>1$, we are measuring their structural parameters at rest-frame optical wavelengths, which should be rather consistent with the $r$-band structural measurements of GAMA galaxies.

We make the following selection cuts to ensure strong completeness and reliability of structural parameters. The $r$-band target selection limits in GAMA are $r<19.0$ for two fields and $r<19.4$ for the third field. We require stellar mass $M_*>10^{10}M_{\odot}$ and $r$-band GALFIT quality flag = 0 (good fits only). Roughly $15\%$ of all galaxies did not satisfy our GALFIT selection criterion, and we verified that these galaxies did not occupy a special region of the sSFR-M$_*$ or UVJ diagrams. We do not split our GAMA sample into finer redshift or stellar mass bins for this study. Our GAMA redshift slice is restricted to $0.005<z<0.12$; the lower limit helps prevent contamination from foreground stars, and the higher limit helps us avoid completeness issues \citep[in combination with our stellar mass cut;][]{taylor11,taylor15}. 

We derive completeness correction weights for every GAMA galaxy that satisfies our selection cuts using the $V_{\rm survey}/V_{\rm max}$ method \citep{schmidt68}. As expected from our selection cuts and the high spectroscopic completeness of GAMA, only $\sim3$\% of the GAMA galaxies that make it past our selection cuts have completeness correction weights $V_{\rm survey}/V_{\rm max}>1$, with the max value being $\sim35$. This confirms that our selection cuts are sufficient to make our sample complete down to $M_*=10^{10}M_{\odot}$, and that we do not actually have to apply completeness correction weights to our measurements (just as with CANDELS).

\section{Semi-analytic Model}\label{sec:samdesc}
One of the main strengths of our study is that we will simultaneously and self-consistently analyze a semi-analytic model (SAM) of galaxy formation in the same way as the observations. This will allow us to track the physical drivers behind galaxy transformations, in terms of both structure and star formation, and explore the many possible physical origins of the transition population in a cosmological context. Here we review only the most salient points of the ``Santa Cruz'' SAM used in this study.  We refer the reader to the following sequence of papers for much greater detail about the physical prescriptions implemented in the SAM, and about the origin and evolution of the SAM itself: \citet{somerville99}, \citet{somerville01}, \citet{somerville08}, \citet{somerville12}, and \citet{porter14}. \citet{brennan15} and \citet{brennan17} also have more in-depth and very relevant discussions about the SAM. Finally, we also recommend the recent review article by \citet{somervillearaa} which discusses SAMs and cosmological hydrodynamical simulations along with a general overview of physical models of galaxy formation and evolution.

We use the mock catalogs that were created for the CANDELS survey (Somerville et al., in preparation). These include lightcones that emulate the geometry of the five CANDELS fields, where the masses and positions of the root halos were drawn from the Bolshoi-Planck N-body simulations \citep{rodriguezpuebla16b}. The Bolshoi-Planck simulations adopt cosmological parameters that are consistent with the Planck constraints \citep{planck14}; $\Omega_{m}$=0.307, $\Omega_{\Lambda}$=0.693, $h$=0.678, with a baryon fraction of 0.1578. Merger trees for each halo in the light cone are constructed using the method of \citet{somerville99b}, updated as described in \citet{somerville08}. We combine the five SAM mock catalogs corresponding to the five different CANDELS fields to achieve excellent number statistics, but we note that each SAM mock catalog is in general much larger than the corresponding observed CANDELS field. Since the CANDELS lightcones represent a very small volume at low redshift, we instead use a $z\sim 0.1$ snapshot drawn from Bolshoi-Planck for our lowest redshift slice.

As halos grow due to gravitational collapse, baryons are accreted into the halo. A standard spherically symmetric cooling flow model is adopted to track the rate at which gas can cool and collapse into the central galaxy \citep[see][for details]{somerville08}. Gas that has cooled and collapsed into a disk is considered available for star formation. The SAM has two prescriptions for star formation. The first prescription is applicable to isolated disks and adopts the Schmidt-Kennicutt relation \citep{kennicuttschmidt98} whereby only gas above a certain critical surface mass density can collapse to form new stars. The second prescription applies to starbursts and is triggered after a merger or an internal disk instability (see below). The efficiency and timescale of a starburst induced by a merger depends on the gas fraction and mass ratio of the progenitors \citep[e.g., see][]{hopkins09}. We note that stars formed during a merger-induced starburst are added to the spheroidal component of the remnant galaxy. SFR estimates in the SAM have been averaged over 100 Myr to replicate the timescales probed by our CANDELS UV+IR based SFRs.

The SAM includes feedback from photoionization, stars and supernovae, as well as active black holes. Photoionization feedback is important only at mass scales much lower than the ones we consider in this paper. In low mass galaxies ($M_*\lesssim10^{10}\,\msun$), the mechanical and radiative feedback from supernovae and massive stars is primarily responsible for outflows of cold gas. Only some fraction \citep[dependent on the halo circular velocity;][]{somerville08} of the outflows are deposited into the hot gas reservoir of the galaxy's halo and allowed to cool again (thereby allowing for future gas inflows), while the rest is driven out of the halo completely, and falls back on a longer timescale. We note that each generation of stars produces heavy elements, which are also ejected from galaxies and deposited either in the ISM, hot halo, or intergalactic medium.

Feedback from AGN is very important in determining the properties of massive galaxies in these SAMs. Seed black holes are initially added to galaxies according to the prescription described in \citet{hirschmann12}. Based on hydrodynamic simulations of galaxy mergers, our SAM assumes that galaxy mergers trigger rapid accretion onto the central BH \citep{hopkins07}. There are two ``modes'' of AGN feedback implemented in our SAM. In the ``radiative mode'' (sometimes called ``bright mode'' or ``quasar mode''), radiatively efficient BH accretion can drive winds that remove cold gas from the galaxy, and eventually shut off the BH growth as well. In addition, hot halo gas can fuel radiatively inefficient BH accretion via Bondi-Hoyle accretion \citep{bondi1952}. This mode is associated with powerful radio jets that can heat the halo gas, suppressing or shutting off cooling. This latter mode is often referred to as ``radio mode'' or ``jet mode'' \citep[see][and references therein]{croton06,sijacki07,somerville08,fontanot09,fabian12,heckman14,somervillearaa}.

We note that mergers and disk instabilities (see below) cause the growth of a bulge and drive gas toward the center where the SMBH lives.  This relationship between bulge growth and AGN activity in the SAM, along with the self-regulated BH growth, leads to final black hole masses and bulge masses that are consistent with the observed $M_{BH}-M_{\rm bulge}$ relation \citep{somerville08,hirschmann12}.

Initially all star formation is assumed to occur in disks. Bulge growth in our SAM occurs through two channels: mergers and disk instabilities.  Mergers directly deposit a fraction of the pre-formed stars from the merging satellites into the bulge component, and also trigger starbursts. The stars formed in these merger-triggered bursts are also deposited in the bulge component. In addition, if the ratio of baryonic material in the disk relative to the mass of the dark matter halo becomes too large, we assume that the disk becomes unstable, and move disk material to the bulge until marginal stability is restored \citep[for more details, see][and references therein]{porter14}. It was shown in \citet{porter14} and in \citep{brennan15} that with our currently adopted recipes, our SAM does not produce enough bulge-dominated galaxies if bulges are allowed to grow only through mergers. We obtain much better agreement with the mass function and fraction of bulge-dominated galaxies when we include the disk instability channel for bulge growth. Note that galaxies that become bulge-dominated through a merger or disk instability can re-grow a new disk and become disk-dominated again through accretion of new gas \citep[see the discussion in][]{brennan15}.

One caveat of the SAM is that morphological transformations are treated as being instantaneous, i.e., following a merger or disk instability the material is added to the bulge in a single timestep. As it is unlikely that morphological transformations act on timescales comparable to the cosmic times spanned by our redshift slices, we do not expect this to significantly affect our results.

We estimate the scale radius of our model disks based on the initial angular momentum of the gas, assuming the gas collapses to form an exponential disk. We include the contraction of the halo due to the self-gravity of the baryons \citep{blumenthal86,flores93,mo98,somerville08b}. The sizes of spheroids formed in either disk instabilities or mergers are estimated using the virial theorem and conservation of energy, including the dissipative effects of gas. Our modeling of spheroid sizes has been calibrated on numerical hydrodynamical simulations of binary galaxy mergers as described in \citet{porter14}, and has been shown in that work to reproduce the observed size evolution of spheroid-dominated galaxies since $z\sim 2$ (see also Somerville et al. in preparation).

The SAM produces a prediction for the joint distribution of ages and metallicities in each galaxy as described in \citet{porter14b}. The predictions are consistent with the observed correlation between age, metallicity and stellar velocity dispersion, and the observed lack of radial trends in age and metallicity, for $z\sim0$ elliptical galaxies \citep[again, see][and references therein]{porter14b}. We combine these age and metallicity predictions with stellar population synthesis models to obtain intrinsic (non-dust-attenuated) stellar energy distributions (SED) which may be convolved with any desired filter response functions. We use the stellar population synthesis models of \citet{bc03} with the Padova 1994 isochrones and a Chabrier IMF. Note that the synthetic SEDs currently do not include nebular emission. We optionally include attenuation of the light due to dust using an approach similar to that described in \citet{somerville12}. For this work, we do not use dust-reddened magnitudes and SFRs from the SAM; instead we correct the observed SFRs for dust reddening, and then compare those de-reddened observed SFRs to the intrinsic SAM SFRs. 

The existence of accurate size estimates for the disk and bulge components in our SAM allows us to do something novel. We can compute composite effective radii and S{\'e}rsic indices using a mapping derived by introducing fake galaxies that are composites of $n=1$ (disk) and $n=4$ (spheroid) components into images and then fitting them with a single S{\'e}rsic profile \citep[see][for details]{lang14,brennan15}. The S{\'e}rsic indices and effective radii that we derive here are light-weighted, in contrast with the stellar mass weighted quantities used in \citet{brennan15}, and should provide a more accurate comparison to the S{\'e}rsic indices and sizes derived from light for our observed sample. However, we note that we do not attempt to include the effects of dust attenuation in our light-weighted quantities. These light-weighted quantities have also been used in \citet{brennan17}, who showed that adopting light-rather than stellar mass-weighted quantities did not qualitatively change their results relative to \citet{brennan15}, but it did result in better agreement between the models and the observations.

\section{Methods}\label{sec:methods}
\subsection{Defining Transition Galaxies}
\label{sec:selection}
We define transition galaxies in a physically motivated way using the sSFR-M$_*$ diagram rather than color-color, color-mass or color-magnitude diagrams \citep[see also][]{brennan15,brennan17}. First, we find the normalization of the SFMS in each redshift slice using the average sSFR of dwarf galaxies with $10^9M_{\odot}<M_{*}<10^{9.5}M_{\odot}$ \citep[since these are known to be overwhelmingly star-forming objects; e.g., see][]{geha12}. We then fit a cubic polynomial to these SFMS normalizations as a function of the age of the Universe, which allows us to easily compute the time evolution of the SFMS normalization. We also derive the linear slope of the SFMS in each redshift slice by calculating the derivative with respect to stellar mass of the average sSFR of galaxies with $M_*\sim10^9M_{\odot}$ and $M_*\sim10^{10}M_{\odot}$. Since allowing the slope to be a free parameter does not significantly change our results, we fix it to zero.

For our six CANDELS redshift slices ($0.5-1.0, 1.0-1.4, 1.4-1.8, 1.8-2.2, 2.2-2.6, 2.6-3.0$), we adopt a conservatively large value of 0.4 dex for the $1\sigma$ observed scatter in the SFMS \citep[this is consistent with the ``intrinsic scatter" of the SFMS measured by][]{kurczynski16}. The SFMS in our GAMA redshift slice ($0.005<z<0.12$) appears to have a larger width of 0.7 dex (clearly evident in the one-dimensional histogram of sSFRs). This might be due to the fact that H$\alpha$ probes SFRs on shorter timescales ($\sim10$ Myr) and thus could be sensitive to larger and more frequent excursions of galaxies below the SFMS \citep[see the relevant GAMA paper by][]{davies16}. The conversion from fiber SFR$_{\rm H\alpha}$ to global galaxy SFR$_{\rm H\alpha}$ likely also introduces an additional $\sim0.15$ dex of scatter \citep{richards16}. We therefore adopt $0.7$ dex for the width of the GAMA SFMS. For the SAM, we adopt 0.4 dex for the width of the SFMS for all redshift slices.

We then define the ``transition region'' to range from $1.5\sigma$ to $3.5\sigma$ below the SFMS (i.e., between 0.6 dex and 1.4 dex below the SFMS). Thus, the offset relative to the SFMS and the width of the transition region are fixed in all redshifts slices. The quiescent region comprises all galaxies further than $3.5\sigma$ (1.4 dex) below the SFMS. Assuming that the observed scatter of the SFMS follows a Gaussian distribution, our upper limit for the transition region of $1.5\sigma$ below the ``mean" (i.e., SFMS normalization) suggests that $>85\%$ of star-forming galaxies would lie above that line, and thus that the contamination in the transition region from scattered SFMS galaxies would be $<15\%$. Similarly, our upper limit for the quiescent region of $3.5\sigma$ below the SFMS normalization suggests that $>99\%$ of star-forming galaxies should lie above that line, and thus that the contamination in the quiescent region from star-forming galaxies would be $<1\%$. Note that these statistical arguments are weaker when applied to the GAMA redshift slice because its SFMS width is larger by 0.3 dex, and therefore its transition region boundaries are shifted down by an additional 0.3 dex as well; this means that the upper and lower boundaries of the GAMA transition region are, respectively, $1.3\sigma$ and $2.4\sigma$ below the SFMS.

\begin{figure*} 
\begin{center}
\includegraphics[width=0.6\hsize]{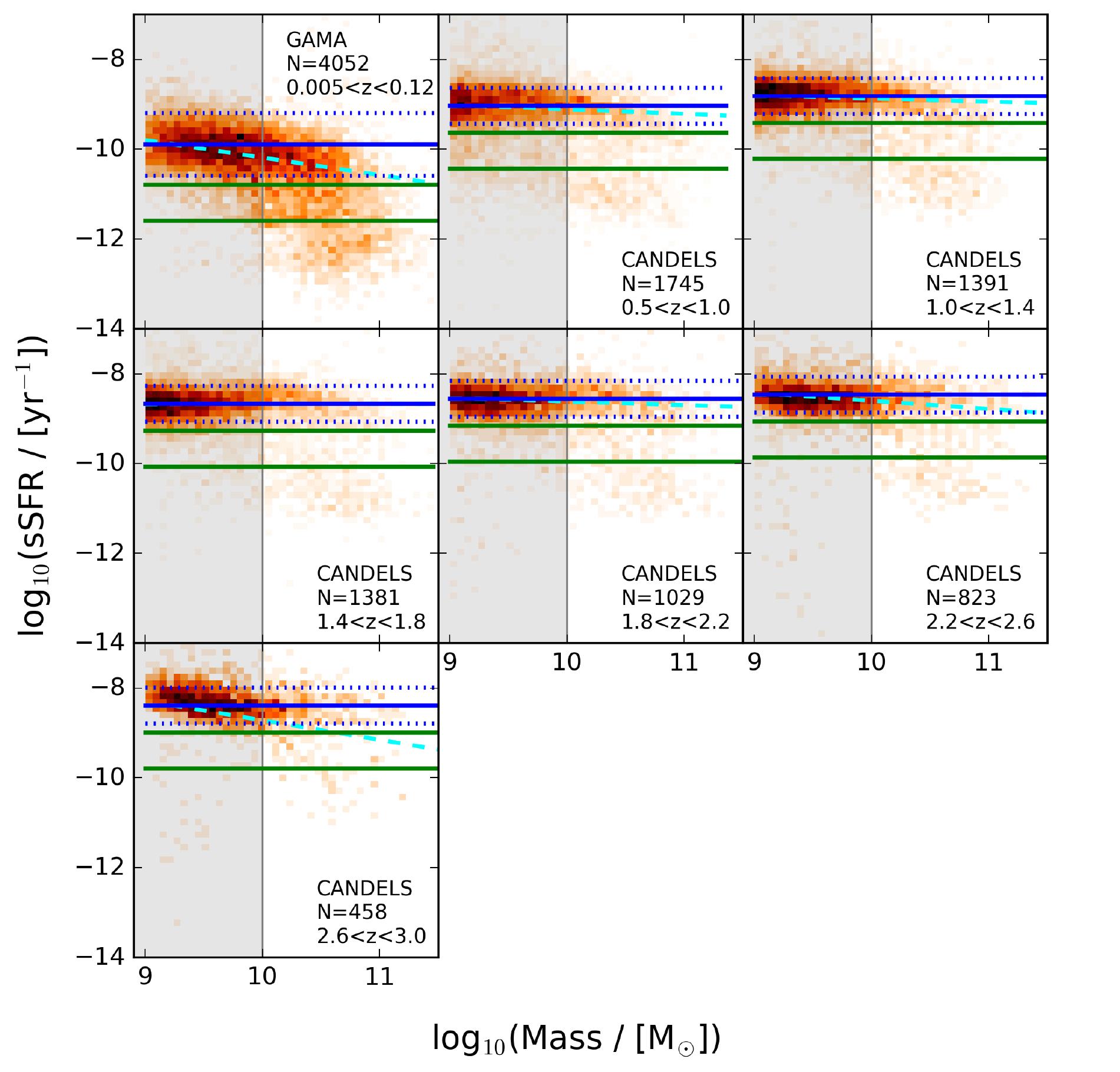}
\includegraphics[width=0.6\hsize]{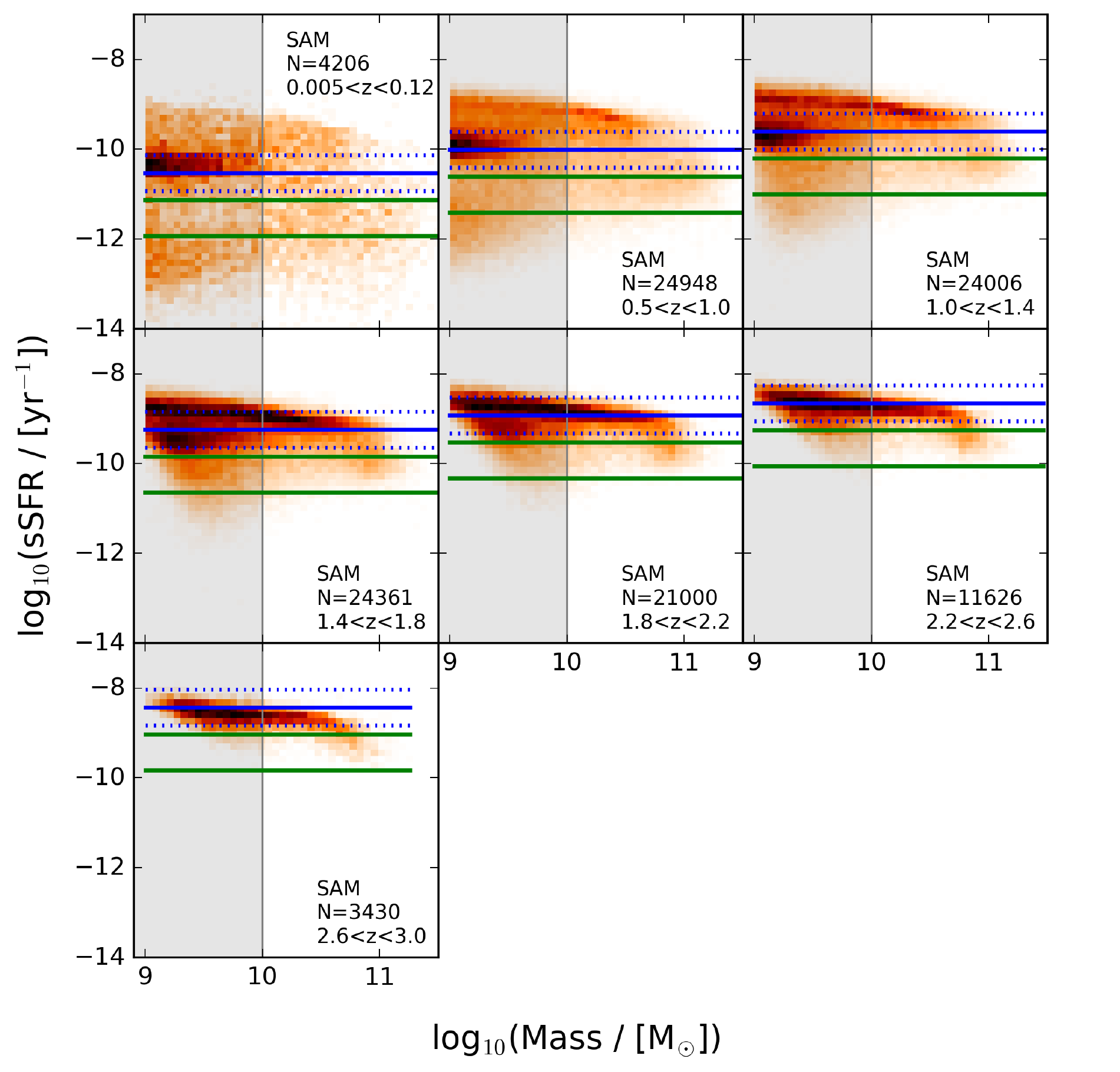}
\end{center}
\caption{Defining transition galaxies in sSFR-M$_*$ space for the observations (top) and for the SAM (bottom). In each panel, the solid blue line shows the SFMS fit (with slope fixed to zero), and the dotted blue lines mark the assumed conservative 0.4 dex $\pm1\sigma$ width of the SFMS (0.7 dex for the $z\sim0.1$ GAMA redshift slice; see text). The two green lines in each panel show the transition region, with the upper line being 0.6 dex ($1.5\sigma$) below the SFMS normalization and the lower line being 1.4 dex ($3.5\sigma$) below the SFMS normalization. The gray shaded region indicates the dwarf galaxy regime where most objects are star-forming; we use this region to self-consistently determine the SFMS normalization (but we do not include these lower mass galaxies in any subsequent analysis). In the observations, the cyan dashed line shows the SFMS fit if the slope is allowed to be a free parameter -- our results and conclusions do not change significantly if we instead define a sloped transition region using the sloped SFMS fits (see \autoref{sec:mmatch} for an extensive discussion of our stellar mass-matching techniques).}
\label{fig:sel}
\end{figure*}

In \autoref{fig:sel}, we show the sSFR-M$_*$ diagram in each of our redshift slices for both the observations and the SAM. We also show the redshift-dependent definition of the transition region for both the observations and the SAM. Our method captures the decreasing normalization of the SFMS toward low redshift. Since we define the transition region in each redshift slice relative to the SFMS in that same redshift slice, the normalization of the transition region also decreases toward low redshift. These features naturally account for the likely possibility that high redshift transition galaxies would be considered star-forming galaxies if they were relocated to $z=0$. 

In the SAM, the SFMS tends to have a lower normalization overall than in the observations; this is known to be a general issue in other models as well \citep[e.g., see the discussions in][]{somervillearaa,dave16}. However, the crucial point is that our method is applied self-consistently and independently to the observations and to the SAM.

\begin{table}
\caption{Coefficients for a cubic polynomial fit to the normalization of the SFMS as a function of the age of the Universe: $\log_{10}(SFMS(z)/\rm yr^{-1})=a_3t^3(z) + a_2t^2(z) + a_1t(z) + a_0$, where $t(z)$ is the age of the Universe at the redshift of interest. These coefficients are valid for $t$ values between roughly $2.5-13$ Gyr. }
\centering
  \begin{tabular}{ccccc}
  \hline
$a_3$ & $a_2$ & $a_1$ & $a_0$ & Sample \\\hline
-0.0011 & 0.0233 & -0.2766 & -7.8597 & GAMA+CANDELS \\
-0.0025 & 0.0787 & -0.8940 & -6.7503 & SAM \\
  \end{tabular}
  \label{tab:sfms}
\end{table}

\subsection{Stellar Mass Dependence}
\label{sec:mmatch}
In our analysis, we attempt to account for the dependence of global galactic structure on stellar mass. This is necessary because more massive galaxies tend to be more bulge-dominated. In each of our redshift slices, the stellar mass distributions of our star-forming, transition, and quiescent galaxies are significantly different from each other. Therefore, it may not be appropriate to compare, e.g., a less massive star-forming galaxy to a more massive transition galaxy since the more massive transition galaxy will naturally have a more prominent bulge. We address this potential stellar mass dependence in three different ways.

Our default approach, which forms the basis for all results shown in this paper, is to perform ``stellar mass-matching" for the transition galaxy subpopulation. Specifically, for each transition galaxy in a given redshift slice, we randomly picked three unique star-forming and three unique quiescent galaxies in the same redshift slice whose stellar masses were within a factor of two of the mass of the transition galaxy. In this way, we constructed ``transition-mass-matched'' samples of star-forming and quiescent galaxies whose structural parameters we could compare to those of transition galaxies. We note that our results are not sensitive to whether or not we apply this stellar mass-matching algorithm. 

As one alternative to our stellar mass-matching approach, we re-did our entire analysis in three stellar mass bins: $10^{10}M_{\odot}<M_*<10^{10.5}M_{\odot}$, $10^{10.5}M_{\odot}<M_*<10^{11}M_{\odot}$, and $M_*>10^{11}M_{\odot}$. Using this mass slice approach, we reproduced the main conclusions of this paper, although there is significantly more scatter in all measurements due to the smaller sample size in each mass bin. 

As a third alternative to our stellar mass-matching algorithm, we re-did our entire analysis by allowing the slope of the SFMS in the sSFR-M$_*$ plane to be a free parameter. Again, our exact quantitative results change slightly, but our conclusions do not.

In the future, it will be important to revisit the non-trivial question of the stellar mass dependence of our results, especially by extending our analysis to lower mass galaxies \citep[see also][]{fang15}. In particular, it will be insightful to consider the relative stellar mass growth of star-forming, transition and quiescent galaxies, assuming that they do indeed form an evolutionary sequence. A naive picture would be that star-forming galaxies should be more massive than their transition and quiescent galaxy descendants, since the latter are forming stars at significantly reduced rates. However, this view is too simplistic because galaxies can grow a significant fraction of their stellar mass through dry mergers and satellite accretion \citep[e.g.,][]{naab07,lackner12b}, and because the most massive objects quench first \citep[e.g.,][]{fontanot09}. A more detailed analysis of the stellar mass growth of individual galaxies as they move between the three different subpopulations is therefore deferred to future work.

\section{Results}\label{sec:results}
\subsection{Structural Distinctiveness and Evolution}
\begin{figure*} 
\begin{center}
\includegraphics[width=0.95\hsize]{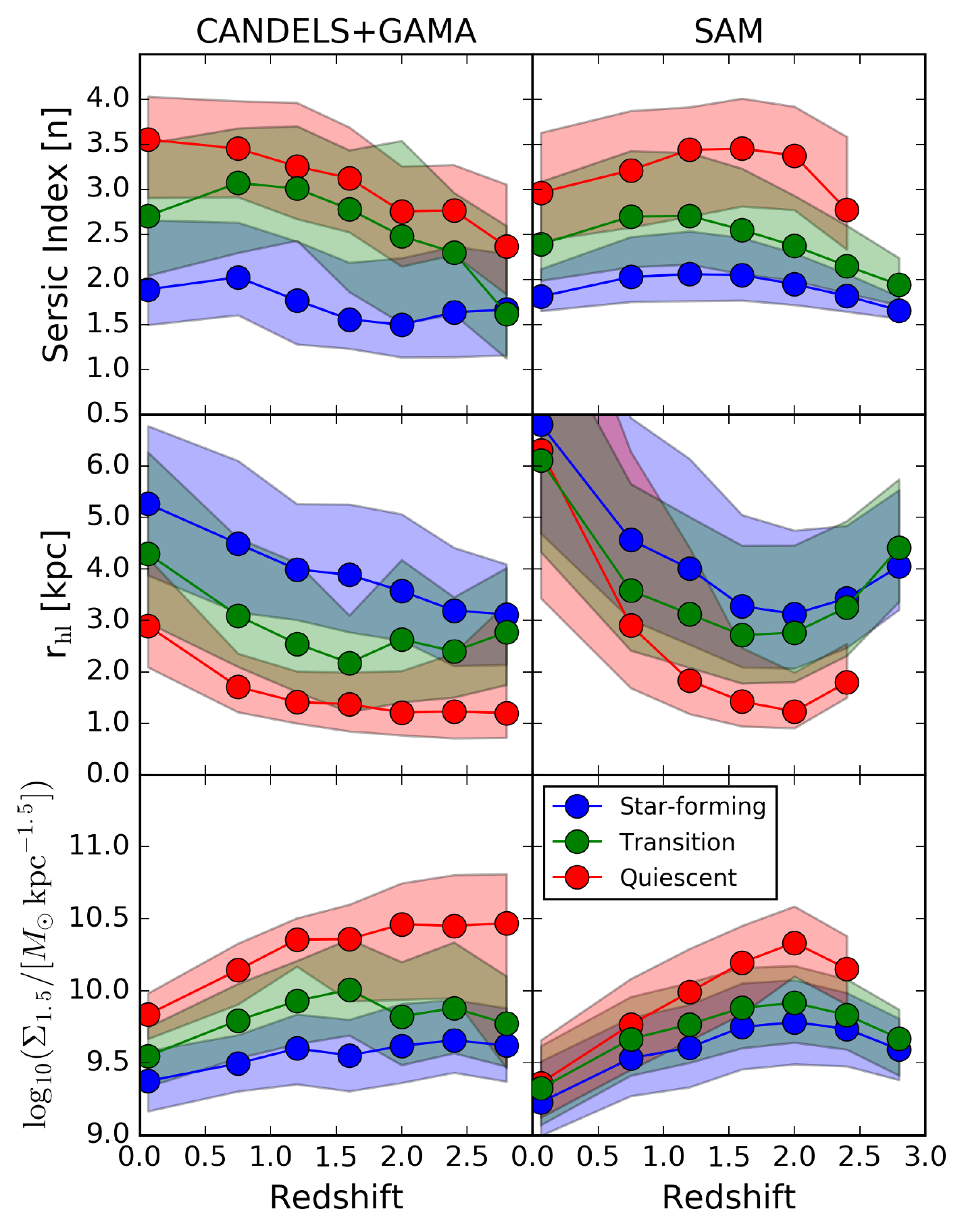}
\end{center}
\caption{The redshift evolution of the S{\'e}rsic index (top row), half-light radius (middle row), and surface stellar mass density (bottom row) for transition galaxies (green), star-forming galaxies (blue), and quiescent galaxies (red). The observations are on the left and the SAM predictions are on the right. The shaded regions span the 25th to 75th percentiles of each distribution, whereas the symbols and lines represent the medians of those distributions. See \autoref{sec:mmatch} for how the star-forming and quiescent subpopulations were stellar mass-matched to the transition subpopulation in each redshift slice. The quiescent predictions from the SAM have been truncated at $z=2.6$ due to the low number of quiescent galaxies in the highest SAM redshift slice. The transition population tends to have intermediate values of these three structural properties relative to the star-forming and quiescent populations in both the observations and the SAM.}
\label{fig:medianz}
\end{figure*}

In \autoref{fig:medianz}, we show the redshift evolution of the S{\'e}rsic index, half-light radius\footnote{Our half-light radii are semi-major axis radii rather than circularized radii ($r_{\rm hl,circ}=\sqrt{q}\times r_{\rm hl}$, where $q\equiv b/a$ is the axis ratio). The latter are more difficult to compare between galaxies since they depend on the shape of each galaxy. However, we also see the same separation between the three subpopulations if we use $r_{\rm hl,circ}$ instead of $r_{\rm hl}$.}, and surface stellar mass density\footnote{The motivation for using $\Sigma_{1.5}\equiv M_*\;r_{\rm hl}^{-1.5}$ is given in \citet{barro13}. In the future, it will be interesting to redo this comparison using the stellar mass density measured within one kpc \citep[i.e., $\Sigma_1$; see][]{cheung12,fang13,barro15}.} for the three subsamples in both the observations and the SAM. In the observations, it is striking how well-separated the median structural properties of the three subsamples are across more than 10 Gyr of cosmic time.  We remind the reader that the results shown here are for stellar mass-matched samples (i.e., we have controlled for stellar mass dependence; see \autoref{sec:mmatch}); it is worthwhile to note that we obtain similar results even without our stellar mass matching algorithm. The cumulative distribution functions (CDFs) underlying \autoref{fig:medianz} as well as the statistical results of two-sample Kolmogorov-Smirnov tests to compare pairs of distributions are given in Appendix \ref{sec:cdf}.

In the observations, we reproduce the well known result that both star-forming and quiescent galaxies have grown in size since $z\sim3$, and that quiescent galaxies are preferentially more compact than star-forming galaxies at all redshifts \cite[e.g.,][]{barro13,vanderwel14,vandokkum15}. What is remarkable, yet also puzzling, is that the transition population seems to remain intermediate between these two populations in terms of S{\'e}rsic index, half-light radius and compactness over this entire interval. 

Intriguingly, we see qualitatively the same trends in the SAM, although with a less pronounced separation between the three populations than what is seen in the observations, especially in the
size and surface stellar mass density. We have confirmed that the separation between the three subpopulations in the SAM, in terms of the S{\'e}rsic index, continues to be seen at all redshifts if we use B/T ratio (either light-weighted or mass-weighted). This suggests that the S{\'e}rsic index separation seen in the SAM is not necessarily driven by our assumed mapping from bulge-disk radii and masses to a composite half-light radius and associated single-component S{\'e}rsic index (as we described in \autoref{sec:samdesc}).

We point out that in our SAM, quiescent galaxies at $z\sim0.1$ tend to have much larger half-light radii than in the observations. While this is an important issue that will be addressed in future work, what is crucial for this paper is not the exact normalization of the size and compactness trends for the SAM, but rather the qualitative separation between the three subpopulations.

\begin{table*}
\scriptsize
\resizebox{\linewidth}{!}{ 
\begin{tabular}{ccccccccccc}\hline
Redshift Slice & $n^{\rm SF}$ & $n^{\rm T}$ & $n^{\rm Q}$ & $r_{\rm hl}^{\rm SF}$ & $r_{\rm hl}^{\rm T}$ & $r_{\rm hl}^{\rm Q}$ & $\Sigma_{1.5}^{\rm SF}$ & $\Sigma_{1.5}^{\rm T}$ & $\Sigma_{1.5}^{\rm Q}$ & Sample \\
$-$ & $-$ & $-$ & $-$ & [kpc] & [kpc] & [kpc] & [log M$_{\odot}$ kpc$^{-1.5}$] & [log M$_{\odot}$ kpc$^{-1.5}$] & [log M$_{\odot}$ kpc$^{-1.5}$] & $-$\\\hline
0.005<z<0.12 & $1.39^{+0.77}_{-0.39}$ & $2.20^{+0.80}_{-0.66}$ & $3.05^{+0.48}_{-0.65}$ & $5.26^{+1.51}_{-1.39}$ & $4.29^{+1.97}_{-1.34}$ & $2.89^{+1.32}_{-0.80}$ & $9.38^{+0.21}_{-0.21}$ & $9.55^{+0.20}_{-0.21}$ & $9.84^{+0.15}_{-0.17}$ & GAMA \\
0.5<z<1.0 & $1.52^{+0.61}_{-0.42}$ & $2.57^{+0.61}_{-0.77}$ & $2.95^{+0.53}_{-0.54}$ & $4.49^{+1.61}_{-1.34}$ & $3.09^{+1.50}_{-0.99}$ & $1.71^{+0.65}_{-0.49}$ & $9.50^{+0.20}_{-0.19}$ & $9.80^{+0.25}_{-0.26}$ & $10.15^{+0.18}_{-0.24}$ & CANDELS \\
1.0<z<1.4 & $1.27^{+0.66}_{-0.49}$ & $2.51^{+0.69}_{-0.58}$ & $2.75^{+0.71}_{-0.58}$ & $3.99^{+1.27}_{-0.98}$ & $2.55^{+1.57}_{-0.93}$ & $1.41^{+0.60}_{-0.41}$ & $9.60^{+0.24}_{-0.25}$ & $9.93^{+0.28}_{-0.30}$ & $10.36^{+0.15}_{-0.19}$ & CANDELS \\
1.4<z<1.8 & $1.06^{+0.63}_{-0.32}$ & $2.28^{+0.65}_{-0.92}$ & $2.62^{+0.57}_{-0.60}$ & $3.88^{+1.37}_{-1.11}$ & $2.18^{+0.92}_{-0.96}$ & $1.37^{+0.62}_{-0.53}$ & $9.55^{+0.25}_{-0.25}$ & $10.01^{+0.35}_{-0.32}$ & $10.36^{+0.24}_{-0.43}$ & CANDELS \\
1.8<z<2.2 & $1.00^{+0.74}_{-0.36}$ & $1.98^{+1.06}_{-0.97}$ & $2.25^{+0.50}_{-0.61}$ & $3.57^{+1.49}_{-0.96}$ & $2.63^{+1.54}_{-1.22}$ & $1.21^{+0.81}_{-0.44}$ & $9.62^{+0.29}_{-0.26}$ & $9.82^{+0.38}_{-0.34}$ & $10.46^{+0.28}_{-0.52}$ & CANDELS \\
2.2<z<2.6 & $1.14^{+0.73}_{-0.50}$ & $1.80^{+0.66}_{-0.68}$ & $2.26^{+0.50}_{-0.49}$ & $3.19^{+1.22}_{-1.07}$ & $2.40^{+1.05}_{-0.89}$ & $1.23^{+1.13}_{-0.52}$ & $9.66^{+0.28}_{-0.22}$ & $9.88^{+0.46}_{-0.31}$ & $10.45^{+0.35}_{-0.50}$ & CANDELS \\
2.6<z<3.0 & $1.16^{+0.62}_{-0.50}$ & $1.12^{+0.98}_{-0.49}$ & $1.86^{+0.69}_{-0.53}$ & $3.12^{+0.97}_{-0.98}$ & $2.77^{+1.25}_{-1.02}$ & $1.20^{+2.13}_{-0.48}$ & $9.62^{+0.26}_{-0.25}$ & $9.77^{+0.33}_{-0.30}$ & $10.47^{+0.34}_{-1.01}$ & CANDELS \\
\end{tabular}}
\caption{Redshift evolution of the structural properties of star-forming, transition and quiescent galaxies in the observations. This table corresponds to what is shown in \autoref{fig:medianz}. Each entry in the table gives the median and the 25th and 75th percentile values relative to that median.}
\label{tab:zevolobs}
\end{table*}

\begin{table*}
\scriptsize
\resizebox{\linewidth}{!}{
\begin{tabular}{ccccccccccc}\hline
Redshift Slice & $n^{\rm SF}$ & $n^{\rm T}$ & $n^{\rm Q}$ & $r_{\rm hl}^{\rm SF}$ & $r_{\rm hl}^{\rm T}$ & $r_{\rm hl}^{\rm Q}$ & $\Sigma_{1.5}^{\rm SF}$ & $\Sigma_{1.5}^{\rm T}$ & $\Sigma_{1.5}^{\rm Q}$ & Sample \\
$-$ & $-$ & $-$ & $-$ & [kpc] & [kpc] & [kpc] & [log M$_{\odot}$ kpc$^{-1.5}$] & [log M$_{\odot}$ kpc$^{-1.5}$] & [log M$_{\odot}$ kpc$^{-1.5}$] & $-$\\\hline
0.005<z<0.12 & $1.31^{+0.31}_{-0.16}$ & $1.89^{+0.70}_{-0.40}$ & $2.46^{+0.67}_{-0.54}$ & $6.81^{+3.51}_{-2.12}$ & $6.11^{+2.75}_{-1.78}$ & $6.30^{+4.59}_{-2.87}$ & $9.23^{+0.28}_{-0.23}$ & $9.32^{+0.29}_{-0.25}$ & $9.36^{+0.30}_{-0.24}$ & SAM \\
0.5<z<1.0 & $1.53^{+0.44}_{-0.28}$ & $2.20^{+0.73}_{-0.56}$ & $2.71^{+0.66}_{-0.64}$ & $4.56^{+2.37}_{-1.55}$ & $3.58^{+2.07}_{-1.16}$ & $2.90^{+3.38}_{-1.21}$ & $9.53^{+0.28}_{-0.26}$ & $9.67^{+0.29}_{-0.25}$ & $9.77^{+0.32}_{-0.31}$ & SAM \\
1.0<z<1.4 & $1.56^{+0.48}_{-0.30}$ & $2.20^{+0.70}_{-0.54}$ & $2.94^{+0.47}_{-0.74}$ & $4.00^{+2.13}_{-1.47}$ & $3.12^{+1.89}_{-1.04}$ & $1.83^{+2.58}_{-0.65}$ & $9.61^{+0.30}_{-0.28}$ & $9.77^{+0.29}_{-0.27}$ & $9.99^{+0.30}_{-0.34}$ & SAM \\
1.4<z<1.8 & $1.55^{+0.41}_{-0.28}$ & $2.05^{+0.68}_{-0.49}$ & $2.95^{+0.56}_{-0.64}$ & $3.27^{+1.77}_{-1.18}$ & $2.72^{+1.74}_{-0.94}$ & $1.43^{+1.04}_{-0.48}$ & $9.75^{+0.30}_{-0.29}$ & $9.88^{+0.28}_{-0.28}$ & $10.20^{+0.26}_{-0.36}$ & SAM \\
1.8<z<2.2 & $1.44^{+0.34}_{-0.23}$ & $1.87^{+0.56}_{-0.39}$ & $2.87^{+0.54}_{-0.60}$ & $3.12^{+1.62}_{-1.06}$ & $2.76^{+1.69}_{-0.96}$ & $1.23^{+0.76}_{-0.32}$ & $9.78^{+0.29}_{-0.29}$ & $9.92^{+0.26}_{-0.27}$ & $10.34^{+0.25}_{-0.23}$ & SAM \\
2.2<z<2.6 & $1.31^{+0.25}_{-0.17}$ & $1.65^{+0.45}_{-0.30}$ & $2.27^{+0.81}_{-0.44}$ & $3.43^{+1.41}_{-0.96}$ & $3.25^{+1.68}_{-0.94}$ & $1.80^{+0.75}_{-0.30}$ & $9.74^{+0.25}_{-0.26}$ & $9.83^{+0.25}_{-0.24}$ & $10.16^{+0.23}_{-0.25}$ & SAM \\
2.6<z<3.0 & $1.15^{+0.15}_{-0.09}$ & $1.44^{+0.30}_{-0.22}$ & $-$ & $4.05^{+1.49}_{-0.84}$ & $4.42^{+1.33}_{-1.05}$ & $-$ & $9.60^{+0.21}_{-0.21}$ & $9.67^{+0.20}_{-0.26}$ & $-$ & SAM \\
\end{tabular}}
\caption{Same as \autoref{tab:zevolobs} but for the SAM. Since there are so few quiescent galaxies in the SAM at $2.6<z<3.0$, we do not measure the distribution of structural properties for quiescent galaxies in that redshift bin.}
\label{tab:zevolsam}
\end{table*}

\subsection{The Transition Fraction Across Cosmic Time}\label{sec:tfrac}
\begin{figure*} 
\begin{center}
\includegraphics[width=\hsize]{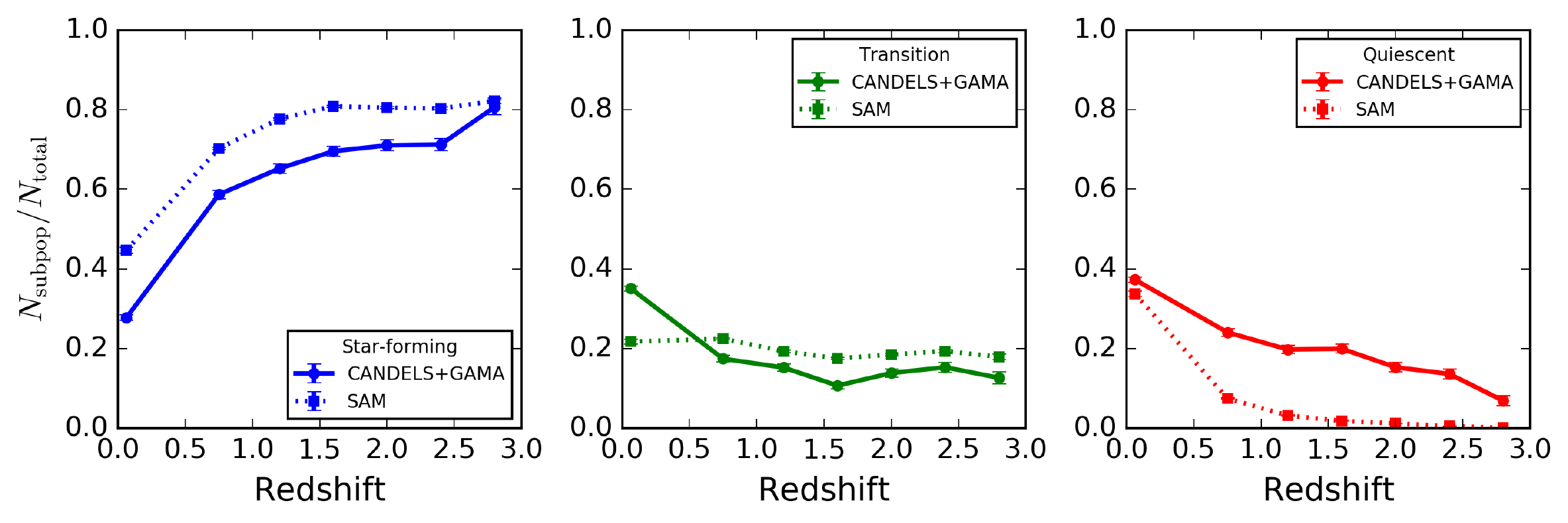}
\end{center}
\caption{The redshift evolution of the fraction of galaxies in the three different subpopulations in the observations (solid lines) and the SAM (dotted lines). Star-forming galaxies are shown in blue (left panel), transition galaxies in green (middle panel), and quiescent galaxies in red (right panel). The errorbars were computed via bootstrapping of the sSFR-M$_*$ diagram. The quiescent fraction builds up toward low redshift while the star-forming fraction decreases -- for both the observations and the SAM. The transition fraction remains roughly constant at all redshifts in the SAM; this is also true for the observations except there is a significant increase from $z\sim0.5$ to $z\sim0.1$. The fact that the SAM quiescent fraction at $z\sim0.1$ agrees well with the quiescent fraction of GAMA suggests that the rate at which galaxies are beginning to quench at high redshift in the SAM is correct. However, the deficit of high-redshift quiescent galaxies suggests that quenching timescales are too long in the SAM.}
\label{fig:tfrac}
\end{figure*}

In \autoref{fig:tfrac}, we show how the fraction of all galaxies that are classified as star-forming, transition and quiescent evolves since $z=3$. The fractions of star-forming, transition, and quiescent galaxies in each redshift slice for the observations and the SAM are respectively given in \autoref{tab:detobs} and \autoref{tab:detsam}. We remind the reader that, in this paper, we are focusing only on massive galaxies with $M_*>10^{10}M_{\odot}$. As has been known for some time \citep[e.g.,][]{bell04,faber07}, the fraction of all massive galaxies that are quiescent has risen considerably. We see in \autoref{fig:tfrac} that this trend is reproduced since $z\sim3$ even when explicitly defining a transition population. Interestingly, the transition fraction is relatively constant between $z=3$ and $z=0.5$ in both the observations and the SAM.

It is immediately apparent from \autoref{fig:tfrac} that at high redshift ($z>0.5$), the SAM underproduces quiescent galaxies. However, the fact that by low redshift the quiescent fraction of the SAM agrees relatively well with that of the observations (down to a discrepancy of $\approx5\%$) suggests that the overall rate at which galaxies begin to quench in the SAM is correct, but that quenching events generally do not happen early enough and that quenching timescales tend to be too slow. We note that if the transition and quiescent populations are grouped into one category (the classical idea of one ``quenched fraction'' for all galaxies below the SFMS), then we would find better agreement with observations, although still with hints that quenching is not efficient enough at high redshift in the SAM \citep[see][and references therein]{brennan15}.

We point out that there is a roughly factor of two increase in the observed transition fraction at very low redshifts, and that no such rapid increase is seen in the SAM. This might be an artifact of our different SFR indicators for GAMA (H$\alpha$, which probes SFRs on 10 Myr timescales) and CANDELS (NUV+MIR, which probes SFRs on 100 Myr timescales). However, it is also entirely possible that the rise is real: our CANDELS observations end at $z=0.5$ and our GAMA observations only go up to $z=0.12$. In the $\sim3.5$ Gyr that have elapsed between these two limiting redshifts, a large number of galaxies could have finally consumed their gas supply and fallen into the transition region. This is also at sufficiently high redshifts that galaxies would still have time to undergo mergers. In the future, it will therefore be interesting to bridge our observational results from the CANDELS and GAMA surveys with observations of ``intermediate-redshift" transition galaxies.

\begin{table*}
\centering
  \begin{tabular}{ccccc}
  \hline
Redshift Slice & f$_{\rm SF}$ & f$_{\rm T}$ & f$_{\rm Q}$ & Sample \\\hline
0.005<z<0.12 & 0.277$\pm$0.007 & 0.350$\pm$0.006 & 0.372$\pm$0.006 & GAMA \\
0.5<z<1.0 & 0.586$\pm$0.011 & 0.174$\pm$0.008 & 0.240$\pm$0.010 & CANDELS \\
1.0<z<1.4 & 0.652$\pm$0.012 & 0.151$\pm$0.009 & 0.197$\pm$0.011 & CANDELS \\
1.4<z<1.8 & 0.695$\pm$0.012 & 0.106$\pm$0.008 & 0.199$\pm$0.011 & CANDELS \\
1.8<z<2.2 & 0.710$\pm$0.014 & 0.138$\pm$0.010 & 0.152$\pm$0.012 & CANDELS \\
2.2<z<2.6 & 0.712$\pm$0.015 & 0.152$\pm$0.012 & 0.135$\pm$0.012 & CANDELS \\
2.6<z<3.0 & 0.806$\pm$0.018 & 0.126$\pm$0.015 & 0.069$\pm$0.012 & CANDELS \\
  \end{tabular}
  \caption{The fraction of star-forming, transition, and quiescent galaxies in each redshift slice in the observations. The errorbars were computed via bootstrapping of the sSFR-M$_*$ diagram.}
  \label{tab:detobs}
\end{table*}

\begin{table*}
\centering
  \begin{tabular}{ccccc}
    \hline
Redshift Slice & f$_{\rm SF}$ & f$_{\rm T}$ & f$_{\rm Q}$ & Sample \\\hline
0.005<z<0.12 & 0.446$\pm$0.008 & 0.217$\pm$0.006 & 0.337$\pm$0.007 & SAM \\
0.5<z<1.0 & 0.702$\pm$0.003 & 0.224$\pm$0.003 & 0.074$\pm$0.002 & SAM \\
1.0<z<1.4 & 0.776$\pm$0.003 & 0.193$\pm$0.003 & 0.031$\pm$0.001 & SAM \\
1.4<z<1.8 & 0.808$\pm$0.003 & 0.174$\pm$0.002 & 0.018$\pm$0.001 & SAM \\
1.8<z<2.2 & 0.804$\pm$0.003 & 0.184$\pm$0.003 & 0.011$\pm$0.001 & SAM \\
2.2<z<2.6 & 0.802$\pm$0.004 & 0.193$\pm$0.004 & 0.005$\pm$0.001 & SAM \\
2.6<z<3.0 & 0.821$\pm$0.007 & 0.178$\pm$0.007 & 0.000$\pm$0.000 & SAM \\
  \end{tabular}
  \caption{The fraction of star-forming, transition, and quiescent galaxies in each redshift slice in the SAM. The errorbars were computed via bootstrapping of the sSFR-M$_*$ diagram.}
  \label{tab:detsam}
\end{table*}

\subsection{Average Population Transition Timescale as a Function of Redshift}\label{sec:transit}
We are now in a unique position to place an upper limit on the average population transition timescale out to $z=3$, by explicitly using the transition population that we have defined in the observations. To do this, we need to make the extreme assumption that transition galaxies observed at any given epoch are all moving from the SFMS toward quiescence, and that they will only make this transition once (i.e., no rejuvenation events or large SFMS oscillatory excursions). In Appendix \ref{sec:numdensity}, we show cubic polynomial fits to the observed number densities of star-forming, transition and quiescent galaxies as a function of redshift. We can use our smooth fits and the redshift-age relation to compute the average population transition timescale as a function of redshift with the following equation:
\begin{equation}\label{eq:transit}
\langle t_{\rm transition}\rangle_{z_1,z_2} = \langle n_{\rm transition}\rangle_{z_1,z_2} \times \left(\frac{d\, n_{\rm quiescent}}{dt}\right)_{z_1,z_2}^{-1}\;.
\end{equation}
Here, $\langle t_{\rm transition}\rangle_{z_1,z_2}$ is the average population transition timescale between two closely spaced redshifts $z_1$ and $z_2$, $\langle n_{\rm transition}\rangle_{z_1,z_2}$ is the average number density of transition galaxies within those two redshifts, and $\left(\frac{d\, n_{\rm quiescent}}{dt}\right)_{z_1,z_2}$ is the change in the number density of quiescent galaxies with respect to the age of the Universe elapsed between those two redshifts. We remind the reader that, in this paper, we focus only on massive galaxies with $M_*>10^{10}M_{\odot}$. 

\begin{figure} 
\begin{center}
\includegraphics[width=\hsize]{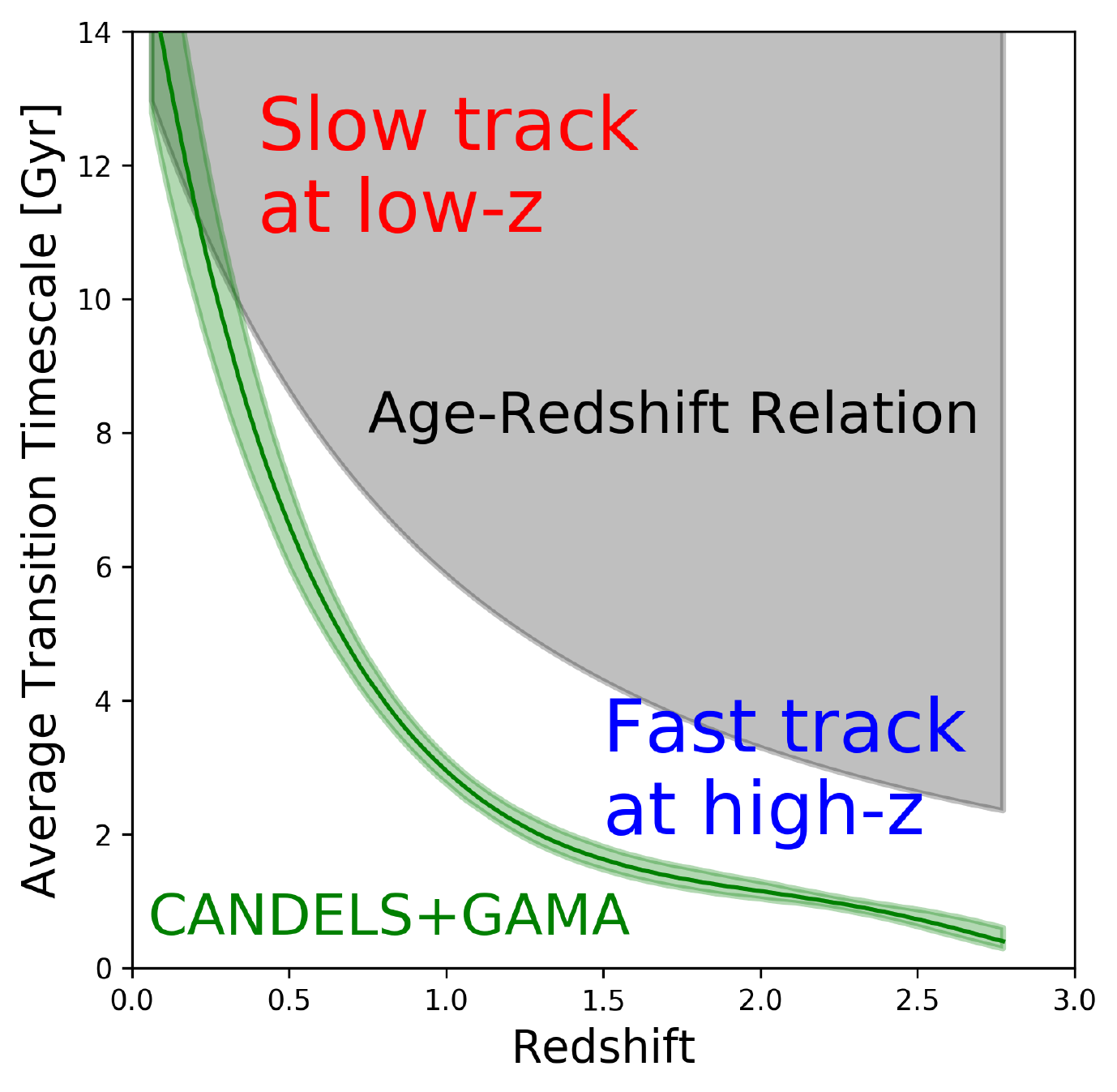}
\end{center}
\caption{Observational upper limit on the average population transition timescale as a function of redshift (solid green line). The green shaded area reflects the $16-84$ percentile uncertainty in our polynomial fits to the number densities of transition and quiescent galaxies as a function of redshift (see Appendix \ref{sec:numdensity}). The gray shaded area reflects transition timescales that would be greater than the age of the Universe at that redshift (i.e., the age-redshift relation). Our measurements are below the age-redshift relation because we do observe quiescent galaxies (even at high-redshift), meaning that they have had enough time at early epochs to make the transition to quiescence. Note how the average population transition timescale is consistent with ``fast track" quenching at high redshift but ``slow track" quenching at low redshift. These calculations are based on massive galaxies only, with $M_*>10^{10}M_{\odot}$.}
\label{fig:transit}
\end{figure}

The results of our calculation are shown in \autoref{fig:transit}. It is immediately apparent that $\langle t_{\rm transition}\rangle_{z_1,z_2}$ rises smoothly from $z=3$ toward $z=0$. This finding explicitly quantifies the notion that, on average, galaxies at high-redshift are on a ``fast track" for quenching ($\sim0.8$ Gyr at $z\sim2.5$) whereas galaxies at low-redshift are on a ``slow track" for quenching ($\sim7$ Gyr at $z\sim0.5$), as schematically described in \citet{barro13}. We point out that our upper limit on the average population transition timescale is below the age-redshift relation, particularly at high redshift. This is a natural consequence of the apparent existence of quiescent galaxies far below the SFMS at these high redshifts, and it suggests that star-forming galaxies are able to make the transition to quiescence faster than the aging of the Universe at these very early times. Note that if galaxies go through the transition region multiple times due to rejuvenation events, then our measurements can also be interpreted as quantifying the average total time spent in the transition region (i.e., the sum of all such transits). 

It is interesting to consider which of the two terms in \autoref{eq:transit} is mainly driving the trend seen in \autoref{fig:transit}. We find that $\langle n_{\rm transition}\rangle_{z_1,z_2}$ is larger than $\left(\frac{d\, n_{\rm quiescent}}{dt}\right)_{z_1,z_2}^{-1}$ at all $z\lesssim2.2$, which means that the change in the number density of quiescent galaxies between any two time steps is smaller than the average number density of transition galaxies within those two time steps. This has at least two possible causes, which are interesting directions for future work but beyond the scope of this paper: (1) a significant fraction of transition galaxies are undergoing slow quenching, SFMS oscillations, or rejuvenation events, and (2) the sSFRs of some transition galaxies might suffer from significant systematic uncertainties due to assumptions made during the SED fitting process, making some of these objects contaminants in the transition region. Regardless of the explanation, we again stress than our result is an \textit{upper limit} on the average population transition timescale as a function of redshift. In the future, it will be interesting to refine \autoref{fig:transit} using smaller redshift and mass slices, as will be afforded by upcoming large surveys of the $z\sim2$ Universe.

\section{Discussion of Observational Results}
\label{sec:discobs}
\subsection{Origin of Structural Distinctiveness}\label{sec:obsinterp}
It is not straightforward to interpret the observational trends seen in \autoref{fig:medianz} because there are many factors that can cause the observed structural distinctiveness of transition galaxies. 
If we assume that the transition population does indeed mostly consist of galaxies moving below the SFMS and toward quiescence (regardless of the timescale), then the range of possibilities is significantly narrowed down. Although naive, such an assumption has at least some basis in our theoretical understanding of galaxy evolution (see our extensive discussion in \autoref{sec:disctheory}) as well as the observational result that the fraction of all massive galaxies that are quiescent is increasing toward low redshift (see again \autoref{fig:tfrac}).

One picture is that galaxies experience ``compaction'' through a dissipative process such as a merger or disk instability. The resulting increase in central stellar density (i.e., bulge growth) is thought to be causally connected with the process that leads to quenching (e.g., AGN feedback). The expected sequence in this picture, in which structural and morphological transformation precedes quenching, seems to lead to a natural explanation of the observed trends. There are also many other findings, both observational and theoretical, that support this ``compaction" picture, at least for high-redshift \citep[e.g., see][]{wuyts12,nelson12,patel13,dekel14,zolotov14,barro15,tacchella15,vandokkum15,tacchella16,tacchella16b,nelson16,huertascompany16}.

It is worthwhile to comment on the two phase formation scenario for early-type galaxies \citep[e.g.,][]{naab07,oser10,lackner12b}. In this scenario, the progenitors of compact quiescent galaxies are formed at high redshift through dissipational processes, and the compact quiescent galaxies themselves then undergo dramatic size evolution toward low redshift via dissipationless processes like dry minor mergers that make them grow far more in size than mass. We see in \autoref{fig:medianz} that quiescent galaxies in both the observations and the SAM get less compact toward low redshift, and that this is also true to a lesser extent for the transition population. \citet{porter14} showed that our SAM can reproduce this trend not just via the commonly assumed dissipationless build-up of the outskirts of high redshift compact quiescent galaxies. Additional physical processes can also act to increase the size and therefore decrease the compactness of quiescent galaxies over cosmic time. These include mixed mergers between disk-dominated and bulge-dominated galaxies, the regrowth of stellar disks in high redshift compact quiescent galaxies, and the decreasing effectiveness of dissipation for producing compact galaxies as the overall gas fraction itself decreases with redshift \citep[see section 5 of][]{porter14}. These latter processes are fundamental for producing galaxies in the SAM that transition between the SFMS and varying ``degrees of quiescence" on a variety of timescales and with a diversity of bulge formation histories (we provide a comprehensive theoretical discussion about the physical origin of transition galaxies in the SAM in \autoref{sec:diversity}).

Furthermore, an important factor to take into account when studying galaxies across such a wide range of redshifts is the concept of ``progenitor bias'' \citep[e.g., see][]{lilly16}. In this scenario, since star-forming galaxies increase in size over time but cease to grow as much after they quench, transition and quiescent galaxies will naturally be more compact than star-forming galaxies observed at the same epoch. In particular, if transition galaxies in a given redshift slice indeed began to quench more recently than quiescent galaxies in the same redshift slice, then we might expect the transition galaxies to be more extended than the quiescent galaxies (after controlling for stellar mass dependence). It is important to note that in this picture, there is no need for ``compaction'' -- transition and quiescent galaxies are more compact than star-forming galaxies not because any mass was transferred toward or grown in their centers, but rather simply because they stopped increasing in size at an earlier epoch, when all galaxies were smaller. 

It seems likely that ``progenitor bias" plays some role in explaining the structural distinctiveness of star-forming, transition, and quiescent galaxies. However, it is still unclear whether it alone can account for all of the observed effect. It is quite possible that both the ``progenitor bias" picture and the ``compaction" picture are at play in the Universe. In our SAM, progenitor bias plays some role but cannot by itself fully reproduce the structural differences between the star-forming, transition and quiescent populations while simultaneously matching other observational constraints \citep[see section 6.2.3 of][]{brennan17}.

\subsection{Using the Transition Population to Probe Systematic Uncertainties}\label{sec:obserr}
While it is indeed promising and compelling that models are beginning to at least qualitatively reproduce observational results derived from statistical samples of galaxies \citep[see][]{somervillearaa}, it is sobering to realize just how many basic questions arise when we try to explicitly define and study this so-called ``transition" population, which we believe must exist in one form or another. A major issue is that we want to study rest-frame colors, relatively ``instantaneous" SFRs, and ultimately the full SFHs of galaxies, but all of these are based on fundamental assumptions made during the SED fitting process (which, in the end, relies critically on getting the redshift correct). If there are any fundamental flaws in our SED fitting assumptions (e.g., universal IMF, universal dust attenuation law, simple SFH parameterizations, assumed light profiles for bulge-disk decompositions, and so on), future attempts at defining and characterizing the transition population may reveal important clues about those problems. Here we briefly comment on potential future improvements to our work.

On the observational side, it will be crucial to obtain a sharper view of \autoref{fig:medianz}, which suggests that bulges directly trace the evolution of galaxies as they fall below the SFMS. The structural measurements that we have used in this paper are based on single-component S{\'e}rsic profile fits \citep{vanderwel12}. Although there are considerable uncertainties associated with bulge-disk decompositions and non-parametric approaches, these are additional tools with which we can observationally probe the relationship between morphological change timescales and transition timescales \citep[e.g.,][]{lackner12,conselice14,bruce14,lang14,peth16,margalef16}. Fitting and comparing structural profiles across the full suite of available multi-wavelength imaging \citep[e.g.,][]{haussler13,vika13}, studying spatial gradients \citep[e.g.,][]{haines15,liu16}, and deriving the full posterior distributions of structural properties of individual galaxies using a Bayesian framework \citep[e.g.,][]{yoon11} may also yield physical insights. In particular, such improvements to structural measurements may allow us to distinguish ``globally quiescent" galaxies from those that are still undergoing star formation outside of the bulge/core component \citep[either inside-out quenching or residual star formation on the outskirts; e.g.,][]{fang12,salim12,wuyts12,nelson12,patel13,abramson14,wellons15,tacchella15,tacchella16b,nelson16,belfiore17}.

On the theoretical side, we have argued that it is better to use SFRs than colors to define the transition population, especially at high redshift. This is because SFRs are relatively ``instantaneous" indicators (10-100 Myr timescales), whereas galaxy colors (depending on the adopted bandpasses) tend to probe the sum of several different stellar populations that may have formed at a variety of redshifts, and can be more sensitive to dust and metallicity (we show where our three subpopulations fall in each CANDELS redshift within the $UVJ$ color-color diagram in \autoref{sec:uvjav}). Nevertheless, SFRs can still be highly uncertain for galaxies that are not actively and continuously forming stars (i.e., galaxies below the SFMS). We know that severe systematic uncertainties in SFRs and stellar masses can arise if underlying assumptions such as a \citet{chabrier} IMF or a \citet{calzetti00} dust reddening law are invalid \citep[e.g., see][]{conroy09,treu10,conroy10a,conroy12,conroy13,cappellari13,reddy15,salmon16}. Uncertainties in the calibration of stellar population synthesis models \citep[e.g.,][]{bc03} and failure to account for the impact of rare but important stellar populations \citep[e.g., thermally-pulsating asymptotic giant branch stars;][]{maraston06,rosenfield14,fumagalli14,villaume15} on galaxy SEDs can also increase systematic uncertainties on observationally-derived physical parameters. These systematic errors are then hard to quantify in large statistical studies that are based on SED fitting, such as ours.

\subsubsection{SFR Uncertainties and the Purple Valley}\label{sec:purple}
It is true that in the observations the width of the SFMS is not merely due to intrinsic scatter alone, but also additional measurement errors. For statistical samples of galaxies such as ours, a detailed uncertainty analysis of SFRs that takes into account our incomplete knowledge of stellar evolution, the IMF, and other topics is often infeasible. If we universally ascribe to each galaxy a conservative SFR measurement error of 0.3 dex (as is often done in studies like ours), then certainly galaxies from one subpopulation can also be consistent with belonging to another subpopulation. For example, star-forming galaxies that may otherwise lie at the intrinsic $1\sigma$ bottom tail of the SFMS (i.e., a distance of 0.4 dex below the SFMS fit) could be scattered further down by an additional 0.3 dex due to measurement errors (so 0.7 dex below the SFMS fit, whereas our transition region spans $0.6-1.4$ dex below the SFMS fit). 

The simple exercise above illustrates that these star-forming galaxies would then also be consistent with a classification as transition galaxies. Although this is a concern, we have defined our transition region to span a wide enough range in sSFRs (0.8 dex) such that not all galaxies could be scattered into or out of it. This idea of galaxies scattering into the transition region is somewhat similar to the idea that the classical green valley might actually be a ``purple valley." The term purple valley was first introduced by \citet{mendez11}, who asked whether the classical green valley might simply be a combination of blue cloud and red sequence galaxies that live in the tails of their parent populations. This includes intrinsically blue or red galaxies that were scattered into the green valley due purely to measurement uncertainties. Could the transition region merely be an analogous combination of intrinsically star-forming and quiescent galaxies that live in the ``tails" of their parent populations? If the SFMS indeed has a physical basis (as we will argue in the next section), then this is unlikely for the following reason. We have effectively defined only two populations: (1) galaxies that are on the SFMS because they have maintained their equilibrium between gas inflows, gas outflows, and star formation, and (2) galaxies that have varying ``degrees of quiescence" below the SFMS, in a continuous sense. As galaxies move further below the SFMS, it becomes less likely that they are maintaining their equilibrium like the average SFMS galaxy; instead, it becomes more likely that they were or are being subject to physical processes that are actively suppressing their star formation. Our view is that the degree of quiescence of galaxies below the SFMS might, in some non-trivial way, reveal clues about the timescales on which their equilibrium was disrupted.

\section{Theoretical Discussion}
\label{sec:disctheory}

\subsection{Physical Significance of Transition Galaxies}\label{sec:physicalsignif}
Our current understanding of galaxy evolution -- based on both observations and theory -- suggests that galaxies flow between the SFMS and varying degrees of quiescence. As is well known, star-forming galaxies occupy a tight sequence in the sSFR-$M_*$ diagram but quiescent galaxies are more diffusely distributed. This is different from classical color-magnitude diagrams, in which it is the quiescent galaxies that form a tight ``red sequence." It is difficult to use this red sequence to theoretically probe the diverse formation histories of quiescent galaxies because: (1) its normalization is due to the physics of stellar evolution, whereby stellar populations approach a maximally red color as they age, and (2) its intrinsic scatter is thought to be due to a degeneracy between age, dust, and metallicity for producing red colors, which has historically been difficult to disentangle both theoretically and observationally. Luckily, the tightness of the SFMS in the sSFR-M$_*$ plane is thought to be due to self-regulation of star formation by stellar-scale feedback processes \citep[e.g.][]{somervillearaa,hopkins14,sparre15,hayward15,tacchella16,rodriguezpuebla16}.\footnote{See \citet{kelson14} for an alternative view about the tight scatter and correlation of the SFMS being due to the central limit theorem. It is still unclear how this interpretation would be affected by the fact that the observed stellar masses of galaxies need not be due entirely to their in situ star formation rates, but that they can also be grown through mergers and accretion of satellites \citep[e.g.,][]{naab07,lackner12b}.} In both sophisticated hydrodynamical simulations and simpler SAMs, galaxies tend to remain close to an ``equilibrium'' condition, in which the net inflow of gas is approximately balanced by outflows and the consumption of gas by star formation \citep[see discussions in][and references therein]{dekelmandelker14,somervillearaa}. When this equilibrium is disrupted by shutting off the inflow of new gas, galaxies naturally drop below the SFMS as they consume their remaining gas \citep[see][for a quenching criterion based on comparing gas depletion and accretion timescales]{tacchella16}.

This highlights how much information transition galaxies potentially carry about the physical cause of the disruption of equilibrium and its timescale. A variety of processes have been suggested in the literature as possible ways to quench galaxies, including virial shock heating of the hot gas halo \citep[sometimes called ``halo quenching'';][]{birnboim03,dekel06}, morphological quenching
\citep{martig09}, tidal and ram pressure stripping of satellites \citep[e.g.,][]{kang08}, radiative and jet mode AGN feedback \citep[see][and references therein]{somervillearaa}, and the general idea of ``compaction" whereby dissipative processes lead to increased central stellar densities and outflows \citep[e.g., ][and our observational discussion in \autoref{sec:obsinterp}]{zolotov14,tacchella16,tacchella16b}. It is worth noting that quenching processes may be ``ejective'' (quenching is caused by removal of the ISM, usually on rapid timescales), ``preventive'' (quenching begins after gas inflows are shutdown and the galaxy consumes its existing gas supply), or ``sterilizing'' (gas remains present in the galaxy, but is rendered unable to form stars efficiently for some reason). These different types of processes should have distinct signatures in terms of the morphology, gas content, and large scale environment of transition galaxies. However, the issue is complicated by the fact that the ``same'' process, broadly construed (e.g., AGN feedback), can manifest in ways that are ejective, preventive, and sterilizing \citep[see][and Brennan et al., in preparation]{choi16}. For example, AGN are known to drive powerful outflows (ejective), cause heating of the extended diffuse gas in halos (preventive), and their hard radiation field may photo-dissociate molecules leading to inefficient star formation (sterilizing).

On the one hand, the qualitative reproduction of the observational trends by the SAM suggests a possible general picture for interpreting the observations. On the other hand, the quantitative discrepancies between the SAM predictions and the observational results may tell us something about the limitations of these models, or revisions that should be made to physical processes within them. The SAMs reproduce the observed trend that quiescent galaxies have the highest, star-forming galaxies have the lowest, and transition galaxies have intermediate S{\'e}rsic index values at all redshifts. In the models, this is a direct result of the connection between the main quenching mechanism (AGN feedback) and the growth of a central bulge \citep[see the extensive discussion in][]{brennan17}. In contrast, the SAM clearly does not produce enough quiescent galaxies at high redshift (\autoref{fig:tfrac}). This is due to some combination of the following factors in the SAM: (1) the overall rate at which star-forming galaxies begin to quench is too low, (2) quenching galaxies take too long to go through the transition region, or (3) quiescent galaxies are rejuvenating too much. We will argue below that the main culprits are that quenching events begin too late and that quenching timescales at high redshift are too long in the SAM.

For simplicity, we will restrict the following discussions only to central galaxies that reach $M_*>10^{10}M_{\odot}$ at $z=0$, and exclude all satellites since they are subject to additional physical processes that we do not focus on in this paper (e.g., tidal stripping). 

\subsection{The Diverse Origin of Transition Galaxies}\label{sec:diversity}
\begin{figure*} 
\begin{center}
\includegraphics[width=\hsize]{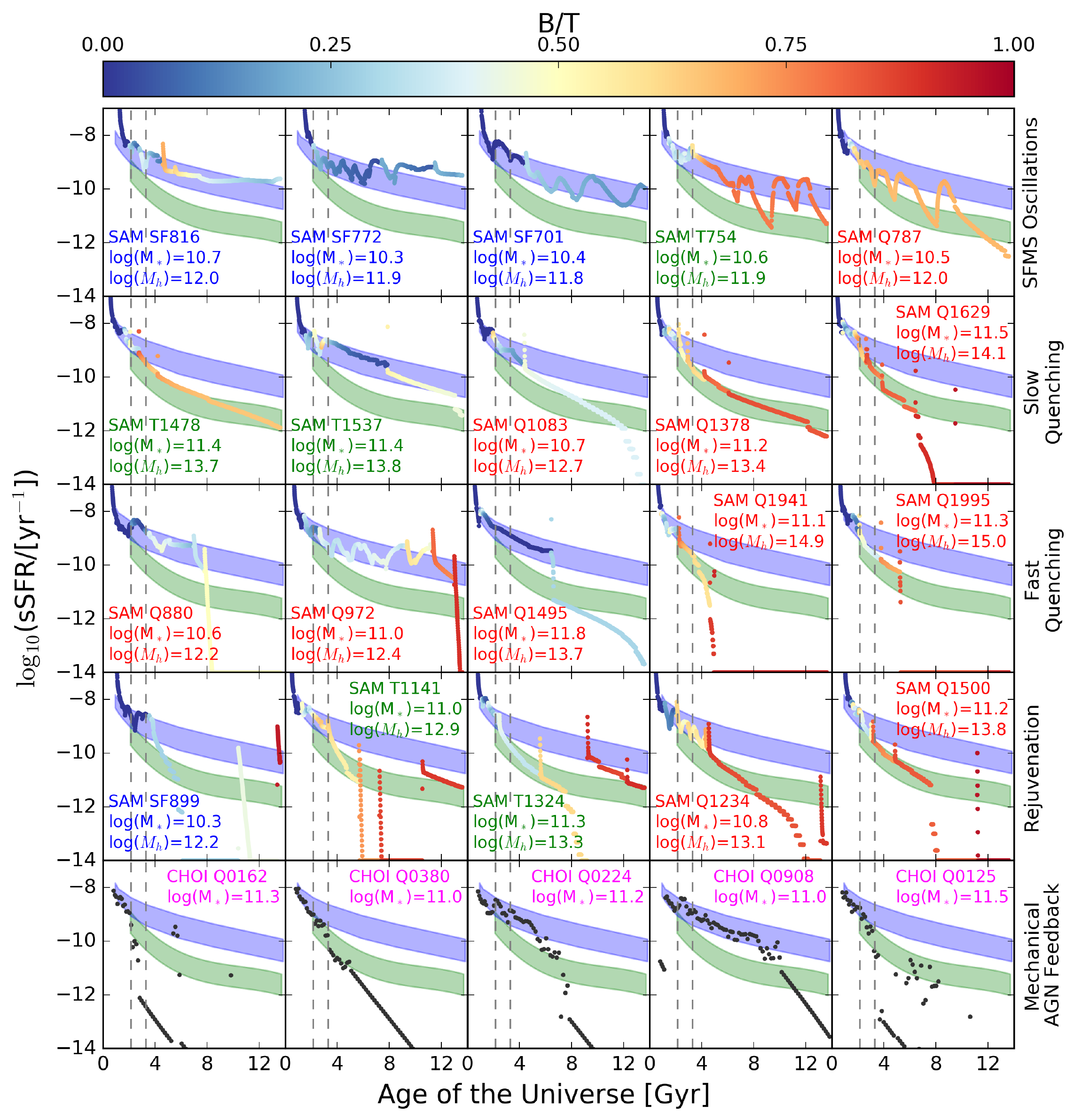}
\end{center}
\caption{Representative SFHs for central galaxies in the SAM, grouped into the four dominant modes of evolution in sSFR-M$_*$ space that we have identified: oscillations on the SFMS (top row), slow quenching (second from top row), fast quenching (middle row), and rejuvenation (second from bottom row). The label for each galaxy is color-coded according to its classification at $z=0$ as star-forming (blue), transition (green), or quiescent (red). Also shown on the bottom row are five representative SFHs of galaxies from the hydrodynamical simulations of \citet{choi16}, which are all quiescent at $z=0$ and include a state-of-the-art implementation of mechanical AGN feedback (magenta labels). Only for the SAM, the colorbar is used to show how the B/T value of the galaxy varies alongside its SFH. The two dashed vertical lines mark $z=3$ (left) and $z=2$ (right). The blue shaded region is the time evolution of the SFMS and its $\pm1\sigma$ scatter from the independent SHARC model \citep{rodriguezpuebla16}, with which the SAM SFMS shows remarkable agreement. The green shaded region is the time evolution of the SAM transition region as defined in this paper. The decreasing normalization of the transition region toward low-redshift accounts for the fact that a high-redshift transition galaxy would be considered a star-forming galaxy if it was relocated to $z=0$. Note the diversity and non-monotonicity of pathways through the transition region and how the effects from stochastic events like mergers and disk instabilities are propagated onto the SFHs and morphological evolutionary histories of SAM galaxies.}
\label{fig:repsfh}
\end{figure*}

Even in population studies, we can learn a lot by first studying the diverse evolutionary histories of individual galaxies \citep[e.g., see][]{brennan15,wellons15,trayford16}. We have qualitatively identified four different physical origin scenarios for transition galaxies based on the diverse SFHs of galaxies in the SAM: oscillations on the SFMS, slow quenching, fast quenching, and rejuvenation. In \autoref{fig:repsfh}, we show twenty representative SFHs from the SAM. We use the colorbar as a third dimension to show how the stellar mass-weighted B/T ratio evolves alongside each SFH. The bottom row of \autoref{fig:repsfh} shows five additional representative SFHs that were pulled from a state-of-the-art hydrodynamical simulation with mechanical AGN feedback \citep{choi16}; these will be discussed in \autoref{sec:momagn}. For reference, in each panel we also show the time evolution of the transition region as defined for the SAM in this paper. The decreasing normalization of the transition region toward low redshift reflects the fact that a galaxy classified as transition at high-redshift would be considered star-forming if it were relocated to $z\sim0$ (based on its sSFR). We also show the time evolution of the SFMS and its $\pm1\sigma$ scatter as predicted by the independent Stellar-Halo Accretion Rate Coevolution model \citep[SHARC;][]{rodriguezpuebla16}, in which the SFR of central galaxies is determined by the overall halo mass accretion rate. The SFMS of the SAM shows remarkable agreement with the SHARC prediction. 

We will now step through the four possible origin scenarios that we have qualitatively identified for transition galaxies in the SAM and discuss their physical causes and implications.

In the first origin scenario, galaxies can undergo oscillations on the SFMS (top row of \autoref{fig:repsfh}). These oscillations are due to variations in a galaxy's gas accretion rate and the interplay between star formation and stellar-scale feedback processes. The overall halo mass accretion rate can also play a role: when the mass accretion rate of a halo drops faster than that of an average halo, the decline in the sSFR of the central galaxy has a steeper slope than the decreasing normalization of the SFMS with redshift (this occurs for halos that assembled their mass earlier than average). In general, these oscillations in the SAM are consistent with the $1\sigma$ scatter of the SFMS \citep[see also the SHARC model;][]{rodriguezpuebla16}. Galaxies tend to remain disk-dominated ($B/T<0.5$) during their oscillations, but this is expected given that they are undergoing rather continuous star formation. \citet{zolotov14} and \citet{tacchella16} found similar oscillatory behavior in their hydrodynamical simulations, and emphasized the importance of ``compaction" events for generating the oscillations (the confinement of the oscillations to the SFMS was due to the interplay between gas depletion and accretion timescales). An intriguing implication of these oscillations is that star-forming galaxies can dip into the transition region briefly and then ascend back onto the SFMS. Two notable examples are shown in \autoref{fig:repsfh}: both T754 and Q787 have quite large excursions and dominant bulges. If such oscillation-induced dips into the transition region are accompanied by significant bulge growth and culminate in quiescence at high redshift, then such galaxies observed during their transition phase may be the so-called ``green nuggets,'' the direct descendants of compact star-forming galaxies and immediate progenitors of compact quiescent galaxies observed at $z\sim2$ \citep{zolotov14,dekel14,tacchella16,barro16}. On the other hand, this first mode can also include rare cases like SF816 and SF772, in which the galaxy has ``lived high'' on the SFMS for its whole life (effectively maintaining a constant SFH since $z\sim3$). It is far above the SFMS at $z=0$ not because it is experiencing a classical starburst, but simply because its halo mass accretion rate (and therefore gas accretion rate) was higher than that of an average SFMS galaxy. 

In the second origin scenario, galaxies undergo ``slow quenching" that can lead to extremely long times spent in the transition region (second from top row in \autoref{fig:repsfh}). This is driven mainly by mergers and the SMBH accretion rate, but is also affected on some level by the halo mass accretion history. We emphasize the diversity of bulge formation histories accompanying this slow quenching pathway: all five representative galaxies shown for this mode in \autoref{fig:repsfh} underwent a merger (which appears to initiate all slow quenching events in the SAM), but not all of them developed a dominant bulge (e.g., Q1083 would be considered a ``disk-dominated" quiescent galaxy at $z=0$). It has long been noted that many galaxies that are observed to live in the classical green valley do not show any obvious signs of recent or ongoing violent star formation suppression mechanisms like radiatively efficient AGN feedback; two prominent examples, at least in terms of the classical green valley, are the Milky Way \citep{licquia15,blandhawthorn16} and M31 \citep{mutch11,williams15}. Resolved stellar population studies of M31 might teach us a lot about this ``slow quenching" mode. \citet{williams15} used the Panchromatic Hubble Andromeda Treasury \citep[PHAT;][]{dalcanton12} to determine that a major global star formation event occurred in M31 roughly $2-4$ Gyr ago. Although the cause of the event is unknown, the main proposed scenarios invoke tidal interactions with M32 and/or M33, or a major merger with another galaxy that became part of what we now call M31 \citep[see section 4 of][and references therein]{williams15}. Detailed bulge-disk-halo-nucleus decompositions of M31's light reveal complex structures even though on a global scale the galaxy would be considered merely ``disk-dominated"; namely, that a massive bulge dominates within $\sim1.5$ kpc, and that the stellar halo exhibits intricate streams \citep[][and references therein]{courteau11,dorman13}. The observations make clear that whatever happened to M31, a rapid quenching event did not simultaneously drive the entire galaxy toward a state of heavy quiescence and heavy ``bulge dominance" --  regardless of whether the SMBH was fed or whether the bulge grew in mass and size.

The third origin scenario for transition galaxies in the SAM requires rapidly quenching galaxies with radiatively efficient AGN feedback that is triggered by mergers (middle row in \autoref{fig:repsfh}). Recall that in the SAM shown here, there are two modes of AGN feedback: (1) radiation pressure-driven winds that correspond to the rapid accretion phase of the black hole and that quickly remove the cold gas, and (2) jet mode feedback that can act as a ``maintenance mode'' and prevent hot halo gas from further cooling and accreting into the galaxy. The significant bulge growth associated with major mergers and the subsequent ``ejective" feedback associated with radiative mode are crucial in our SAM for producing the bulge-dominated quiescent population that we observe at $z>2$ (only 4 Gyr since the Big Bang). However, \autoref{fig:repsfh} reveals that not all fast quenching events in the SAM act at early times (e.g., Q972), and that not all such events lead to ``pure bulge" (elliptical) remnants (e.g., Q880 and Q1495 have roughly intermediate B/T for roughly half the age of the Universe). Furthermore, some ``fast quenching" pathways like Q1941 take $\sim2$ Gyr to get through the transition region at $z\sim3$, but that is a very large fraction of the age of the Universe at those early times.

Finally, in some cases, the ``maintenance mode" of AGN feedback fails to fully do its job of keeping the halo gas hot, and so the gas manages to cool and reignite star formation in the galaxy. Mergers can also bring in gas, causing residual star formation events before the galaxy drops into quiescence once again. This leads to the fourth origin scenario for transition galaxies in the SAM: rejuvenation (second from bottom row in \autoref{fig:repsfh}). It is important to clarify that rejuvenation consists of two phases that typically occur on very different timescales. First, the galaxy's sSFR jumps many orders of magnitude from quiescence back onto the SFMS (or perhaps into the transition region). Then, the galaxy will at some point begin to ``re-fade"; this can certainly be sped up with quenching events as described above. The actual rejuvenation event occurs on a much faster timescale than the subsequent ``re-fading" phase. The reason for this is that a galaxy's color will become bluer due to the appearance of newly-born young stars on a much faster timescale compared to the subsequent reddening of the stellar population. This means that, in general, rejuvenated galaxies in the transition region will be caught during their declining sSFR phase rather than their increasing sSFR phase. With that said, it might still be interesting to speculate about the possible existence of ``slow rejuvenation" tracks (e.g., due to a steady sequence of very minor star formation episodes). Constraining the physical mechanisms that could give rise to such ``slow rejuvenation" tracks, and identifying their corresponding observables, might help place firm limits on the fraction of galaxies in the transition region that are actively rejuvenating rather than once again moving toward quiescence.

All of this begs the question: how do we observationally identify which galaxies in the transition region are merely undergoing large oscillations on the SFMS, slowly fading towards quiescence, rapidly being quenched, or experiencing rejuvenation? In addition to the costly method (especially at high redshift) of constructing non-parametric SFHs using spectroscopy, one way to proceed might be to attempt to link the four evolutionary modes for SFHs that we have identified in the SAM to the many other histories of galaxies (structural, dynamical, and so on). Another way is to explore the predicted range of transition timescales for each of the four modes and their associated physical processes. Ultimately, we would want to understand the relative frequency with which each of the four evolutionary modes occurs in a cosmological context (i.e., their dependence on redshift, stellar mass, halo mass, and so on). Many of these questions are beyond the scope of this paper, but in the remaining subsections, we will briefly explore these topics.

\subsection{Transition Region Occupation Timescales}\label{sec:tocc}
In the previous subsection, we showed representative SFHs and qualitatively identified four origin scenarios for transition galaxies in the SAM. Now we will use the transition population to: (1) understand why the transition fraction is constant as a function of redshift in the SAM, and (2) diagnose why the SAM is under-predicting the quiescent fraction at high-redshift relative to the observations (see again \autoref{fig:tfrac}). Specifically, for galaxies classified as star-forming, transition or quiescent at $z=0$, we can study how much time they have collectively spent in the SFMS, transition region and quiescent region since $z=3$. To do this, we simply trace each galaxy's SFH back to $z=3$ and count up the total time that it has spent in the three different regions of the sSFR-M$_*$ diagram (while accounting for the fact that the typical sSFRs of these three subtypes increase smoothly toward high redshift).\footnote{When studying the evolution of an individual galaxy in the SAM, we follow the main (i.e., most massive) branch of the merger tree.} It is crucial that we do not assume some parameterization for the SFH (e.g., exponentially decaying single-$\tau$ models) since many such parameterizations may not accurately capture the bursty, stochastic, and non-monotonic nature of our SFHs \citep[see also][]{pacifici15}. 

We identify galaxies with rejuvenating SFHs by applying a crude threshold of at most five timesteps since $z=3$ in which: (1) a $z=0$ star-forming galaxy can be quiescent in its history, (2) a $z=0$ transition galaxy can be quiescent in its history, and (3) a $z=0$ quiescent galaxy can be star-forming or transition after the first time in its history that it became quiescent. We find that $13\%$ of $z=0$ star-forming galaxies, $25\%$ of $z=0$ transition galaxies, and $31\%$ of $z=0$ quiescent galaxies in the SAM have experienced rejuvenation events since $z=3$. Because the time spent on the upward rejuvenation track is typically far shorter than the time spent on the subsequent re-fading track, we do not expect the upward portion of rejuvenation tracks to significantly contaminate the transition region occupation timescale distribution. However, since the downward portion of rejuvenation tracks (i.e., the ``re-fading" phase) can significantly increase a galaxy's total time spent in the transition region, we restrict the following analysis to galaxies that have non-rejuvenating SFHs.

In \autoref{fig:occtime}, we show cumulative distribution functions of the SFMS occupation timescale, transition region occupation timescale, and quiescent region occupation timescale. Each of these CDFs is split into three categories based on the classifications of SAM galaxies at $z=0$ as star-forming, transition or quiescent. The CDFs extend up to a maximum timescale of 12 Gyr (roughly the time elapsed since $z=3$). We will now discuss each of the three CDFs in turn.

\subsubsection{Total Time on the SFMS}
Galaxies classified as star-forming at $z=0$ have an average SFMS occupation timescale of $\sim11$ Gyr since $z=3$, which implies that $z=0$ star-forming galaxies have led rather quiet lives in terms of their SFHs (the vast majority of them never left the SFMS). The very small tail toward lower SFMS occupation timescales identifies galaxies that underwent significant oscillations into the transition region, which would naturally lower their total time spent on the SFMS since $z=3$. Of course, the tail toward lower SFMS occupation timescales would become more significant if we included rejuvenating SFHs, but we found above that only a minority ($\sim13\%$) of $z=0$ star-forming galaxies in the SAM have experienced such rejuvenation events since $z=3$. 

Galaxies classified as transition at $z=0$ have an average SFMS occupation timescale of $\sim9$ Gyr since $z=3$. Such a long time spent on the SFMS since $z=3$ could be the result of either SFMS oscillations, a fast quenching event at late times, or a slow quenching event at intermediate/late times. There is a sharp drop-off toward lower values of the SFMS occupation timescale distribution, with effectively no $z=0$ transition galaxies having spent less than 5 Gyr on the SFMS since $z=3$. Such low times spent on the SFMS since $z=3$ would result from galaxies that began to quench at early times. In the SAM, it is rare for a galaxy to still be in the transition region at $z=0$ if it began quenching at $z\sim3$. Most of these galaxies will in fact become quiescent by $z=0$. A spectacular exception in \autoref{fig:repsfh} is T1478, which spent $\sim2$ Gyr since $z=3$ on the SFMS, underwent slow quenching for the next $\sim9$ Gyr, and still remained in the transition region at $z=0$ (interestingly, with an intermediate B/T). 

Galaxies classified as quiescent at $z=0$ spent an average time of $\sim5$ Gyr since $z=3$ on the SFMS. Since we are only considering non-rejuvenating SFHs, this means that the average $z=0$ quiescent galaxy first began to quench no later than $z\sim0.7$ (the redshift corresponding to 5 Gyr after $z=3$). Only $\sim20\%$ of $z=0$ quiescent galaxies began to quench at $z>2$ (these are the ones that spent $<2$ Gyr on the SFMS since $z=3$). A rather large $\sim20\%$ of $z=0$ quiescent galaxies did not begin their quenching event until $z<0.5$ (these galaxies spent $>7$ Gyr on the SFMS since $z=3$ and finished their quenching within the remaining $\sim5$ Gyr).

\subsubsection{Total Time in Transition and Quiescence}
We can gain more physical insight by looking at the CDF of transition region occupation timescales (middle panel of \autoref{fig:occtime}). The most salient feature here is that a large fraction of $z=0$ transition and quiescent galaxies have spent several Gyr since $z=3$ in the transition region, with the average being $\sim2$ Gyr. In the previous section, we found that the average $z=0$ quiescent galaxy first began to quench no later than $z\sim0.7$. Combined with the $\sim2$ Gyr average transition timescale, this means that the average $z=0$ quiescent galaxy first joined the quiescent population by $z\sim0.4$. This explains the rapid upturn in the SAM's quiescent fraction starting at $z\sim0.7$ shown in \autoref{fig:tfrac}. 

To complete the circle, we can include rejuvenating SFHs and ask: how long do SAM galaxies actually remain quiescent after their quenching is first complete? We found above that $31\%$ of quiescent galaxies in the SAM have experienced at least one significant rejuvenation event since $z=3$. We also found that the average $z=0$ quiescent galaxy first joined the quiescent population at $z\sim0.4$ in the SAM. If the average quiescent galaxy did not undergo any rejuvenation, then it should have remained quiescent for $\sim4.5$ Gyr (the time elapsed between $z=0$ and $z=0.4$). This is very close to the actual average time spent in quiescence as shown in the right panel of \autoref{fig:occtime}, where we have included rejuvenating SFHs. This means that, on average, rejuvenated galaxies spend very little time in the SFMS and transition region before rejoining the quiescent population in the SAM. 

Finally, we now remark that the interplay between all of the possibilities discussed above gives rise to the constant transition fraction for the SAM seen in \autoref{fig:tfrac}. In the SAM, this constant transition fraction occurs because galaxies are constantly moving into and out of the transition region on a variety of timescales and from various directions (including, e.g., transition galaxies that undergo a mixed merger and get kicked back up onto the SFMS without ever being able to complete their transition). It is intriguing to wonder whether the constant transition fraction in the observations might also be due to the fact that the transition region is a highway of sorts for galaxy evolution.

\subsubsection{SAM Diagnosis and Comparisons to Other Models}
We know that the overall rate at which galaxies are quenching in the SAM is roughly correct because the SAM quiescent fraction agrees relatively well with the observations at $z\sim0.1$ (see again \autoref{fig:tfrac}). Our analysis above further shows that the SAM has a deficit of quiescent galaxies at $z>0.5$ primarily because quenching events happen too late and quenching timescales are too slow in the SAM. Many of the fast quenching events in the SAM are not beginning early enough or acting quickly enough, so that even the fastest quenching galaxies (e.g., Q1941 in \autoref{fig:repsfh}) are still in the SFMS or the transition region at $z=3$ and do not reach the quiescent region until $z<1$. 

It is imperative to comment on the possibility that the deficit of quiescent galaxies at high redshift in our SAM might also apply to other models, both semi-analytic and hydrodynamic. The extensive study of \citet{lu14} found remarkable agreement between three independent SAMs, one of which was the ``Santa Cruz" SAM used in this paper. Although we did not present a comprehensive study of these other SAMs in this paper, we have verified that a similar issue related to the underproduction of the quiescent population (and even the transition population, unlike for our SAM) at high-redshift exists in at least one more SAM examined by \citet{lu14}. 

On the hydrodynamical side, \citet{trayford16} carried out a comprehensive analysis of the EAGLE cosmological simulation \citep{schaye15}, and found that their $z\sim0$ red galaxies spent a median time of $\sim2$ Gyr in the classical green valley since $z=2$ (see their Figure 10). They interpreted this to mean that their galaxies do not stay in the green valley for long, but their median timescale is not so different from the average transition timescale of our $z=0$ quiescent galaxies ($\sim2$ Gyr since $z=3$). In addition, \citet{feldmann16b} recently found that the ultra-high resolution FIRE simulations \citep{hopkins14} are unable to reproduce the ``reddest" massive quiescent galaxies observed at $z=2$ (based on rest-frame $UVJ$ color-color selection criteria). Similarly, \citet{bluck16} also recently suggested that quenching might not be efficient enough in the Illustris cosmological hydrodynamical simulation \citep{genel14,vogelsberger14}, based partially on an analysis of the $z=0$ transition population in SDSS and Illustris. 

Interestingly, in the hydrodynamical simulations of \citet{cen14}, most galaxies that are in the red sequence at $z=0.62$ (the computational redshift limit of their simulations) spent only $300\pm150$ Myr in the classical green valley. However, they also find that a whopping $40\%$ of their massive galaxies that are in the green valley at $z\sim1$ do not actually become red by $z=0.62$. In other words, even in the promising simulations of \citet{cen14}, there are a startling number of galaxies that linger in the green valley for $\sim2$ Gyr (the age difference between $z=1$ and $z=0.62$), and this timescale would likely only increase if they could extend their simulations down to $z=0$ (an additional 6 Gyr since their computational limit of $z=0.62$). 

Even these few qualitative comparisons between our SAM and other simulations stress the need to ask why this problem is only now beginning to be noticed in high redshift studies. One simple possibility is the splitting of a sample into only two subpopulations of star-forming and ``quiescent" galaxies. In such a scenario, the modeled ``quiescent" fraction can be boosted by including transition galaxies and perhaps also galaxies in the lower tail of the SFMS. This is one reason to adopt our physically and statistically motivated approach described in \autoref{sec:selection}. Two alternative ways forward, instead of explicitly categorizing the transition population as in our paper, are to: (1) check whether simulations reproduce the full observed spread in the ``degree of quiescence" below the tight SFMS, in a continuous sense \citep[see][]{brennan17}, and (2) construct and compare ``sSFR functions" \citep[see][]{dave16b}. 

\subsection{Quenching Timescales in Hydrodynamical Simulations with Mechanical AGN Feedback}
\label{sec:momagn}
Although the SAM includes radiation pressure-driven winds from AGN, the implementation is based on an earlier generation of hydrodynamical simulations. We therefore examine recent high-resolution hydrodynamical simulations presented in \citet{choi16}, which include a more detailed and physical implementation of AGN driven winds. Thirty halos that span $M_{\rm halo}\sim10^{12-13.4}M_{\odot}$ were simulated. In these simulations, both the thermal energy and the momentum arising from radiation pressure in the unresolved broad line region are injected into the gas surrounding the accreting black hole. As shown by \citet{choi15,choi16}, this mechanical feedback from AGN drives powerful galaxy-wide outflows that not only sweep the ISM out of the galaxy, but also shock-heat the surrounding hot gas halo leading to strong quenching over long timescales. 

In the bottom row of \autoref{fig:repsfh}, we show representative SFHs from several ``zoom-in'' simulations of individual massive halos (five out of the full sample of thirty halos). In the few cases where the galaxies continue to form stars below $z=2$ (such as Q0224 and Q0908 in \autoref{fig:repsfh}), the galaxies seem to follow the SHARC SFMS as was the case for the SAM above. Therefore, given the limited number of halos that we have for these simulations, when we trace the SFHs of these galaxies back to $z=3$ and count up the total that they have spent in the SFMS, transition region and quiescent region, we use the boundaries as defined for the SAM.

The representative SFHs from \citet{choi16} shown in \autoref{fig:repsfh} reveal rather abrupt and quick quenching, and once a galaxy becomes quiescent, it tends to stay that way. If one of these quiescent galaxies undergoes rejuvenation, the rejuvenated remnant is more likely to end up in the transition region rather than the SFMS (e.g., Q0125, bottom right of \autoref{fig:repsfh}). This is a natural byproduct of the fact that mechanical AGN feedback acts not only ``ejectively," but also ``preventively" as described in \autoref{sec:physicalsignif}. All of the above comes together nicely to reproduce the giant elliptical galaxies that we observe at $z\sim0$, which are quiescent in every sense of the word (i.e., truly ``red and dead").

\begin{figure*} 
\begin{center}
\includegraphics[width=\hsize]{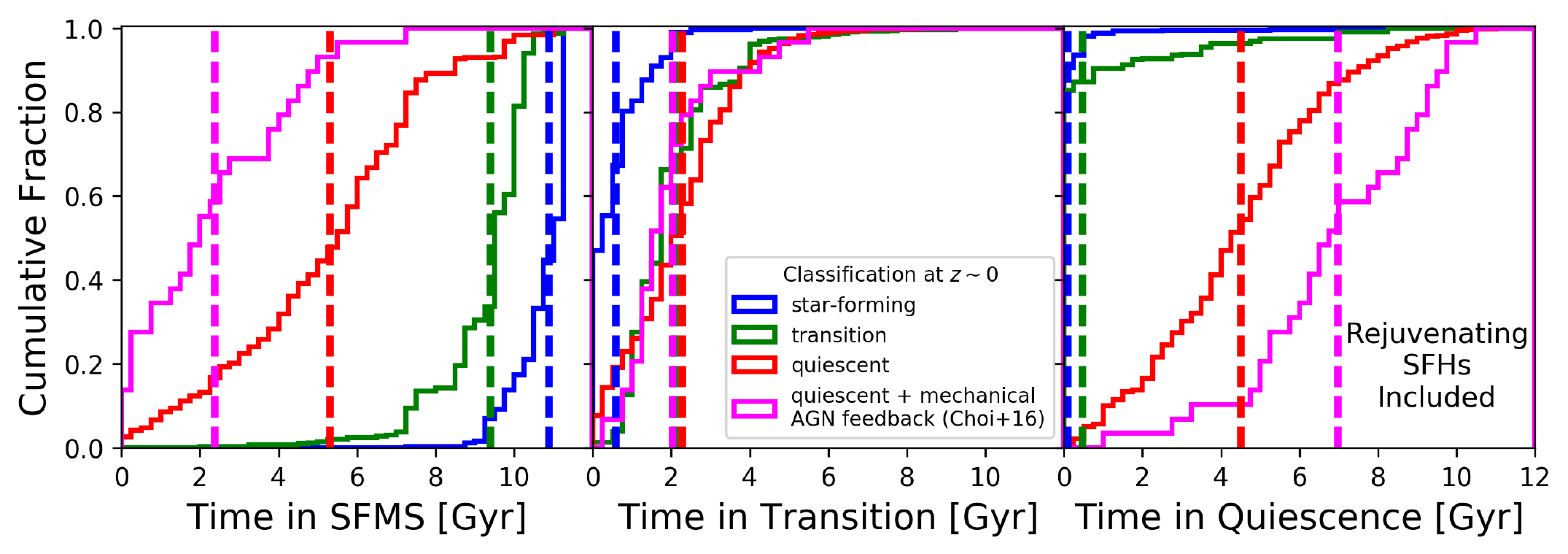}
\end{center}
\caption{Normalized cumulative distribution functions for how long galaxies have spent in the SFMS, transition region, or quiescent region since $z=3$. The CDFs are split into three separate ones for galaxies that are classified at $z=0$ as star-forming (blue), transition (green), or quiescent (red). Also plotted are the occupation timescales for galaxies in the hydrodynamic simulations of \citet{choi16}, which are all classified as quiescent at $z\sim0$ (magenta). In each panel, the median corresponding to each CDF is shown by a vertical dashed line. For the SFMS and transition region occupation timescales, we only use non-rejuvenating SFHs (as described in \autoref{sec:tocc}). However, for the quiescence timescale distributions, we include rejuvenating SFHs to better appreciate how few $z=0$ star-forming and transition galaxies have undergone a rejuvenation event since $z=3$. Note how a non-negligible fraction of $z=0$ quiescent and transition galaxies in the SAM have spent $>2$ Gyr in the transition region since $z=3$.}
\label{fig:occtime}
\end{figure*}

However, \autoref{fig:occtime} reveals that the average time spent on the SFMS by the galaxies in these hydrodynamical simulations since $z=3$ is $\sim2$ Gyr, which means that on average the galaxies of \citet{choi16} do not join the quiescent region until $z<2$. This was also the fundamental problem in the SAM (and in other hydrodynamical simulations, as mentioned in the references above): not even the fastest quenching events at high redshift act quickly enough, so that many fast-quenching galaxies are still in the SFMS or transition region at $z\sim3$. In the hydrodynamical simulations of \citet{choi16}, Q0162 (bottom left of \autoref{fig:repsfh}) is the earliest quenched galaxy: it was already nearly in the quiescent region by $z=3$. But, Q0162 is in a class of its own among the sample of thirty halos from \citet{choi16}; the remaining galaxies take even longer to quench.

Our finding suggests that this more sophisticated treatment of mechanical AGN feedback is a promising way to produce realistic local giant elliptical galaxies and that it can help boost the quiescent fraction in the SAM at $z<2$. However, it is not yet sufficiently clear whether this implementation alone can solve the deficit of quiescent galaxies in the SAM at even higher redshifts ($z\sim3$). In order to produce heavily quiescent galaxies by $z\sim3$, it might be necessary to begin the quenching process at $z\gg3$. One possibility might involve coupling the mechanical feedback from growing SMBHs to the stronger effects expected from clustered supernovae \citep[e.g.,][]{gentry16}. It is also worth mentioning that the simulations of \citet{choi16} span a limited halo mass range ($M_{\rm halo}\sim10^{12-13.4}M_{\odot}$), and that the very high redshift quiescent population might represent the progenitors of even higher halo mass galaxies. Finally, we caution the reader that there are systematic uncertainties in the SFRs of observed high-redshift galaxies that have not yet been thoroughly explored (see again our discussion in \autoref{sec:obserr}).

\subsection{Morphological Dependence of Quenching Timescales}\label{sec:transitbt}
It is natural to ask whether there is any clear-cut dependence of transition timescales on the final morphology of a quiescent galaxy in the SAM. This question likely depends on stellar mass and halo mass (at least), but we can still carry out a general theoretical test of the two extreme morphology-dependent scenarios proposed by \citet{schawinski14}: (1) low-redshift quiescent disk-dominated galaxies were preferentially subject to slow quenching mechanisms that preserved their stellar disks, and (2) low-redshift quiescent bulge-dominated galaxies were preferentially subject to fast quenching mechanisms that also rapidly grew their bulges. We might therefore expect that $z=0$ disk-dominated quiescent galaxies preferentially spent much longer times in the transition region compared to $z=0$ bulge-dominated quiescent galaxies. However, it is already obvious from \autoref{fig:repsfh} that, in the SAM, not all slow quenching events result in a disk-dominated galaxy (e.g., Q1629 has $B/T\approx1$), and that not all fast quenching events result in a bulge-dominated galaxy (e.g., Q1495 has $B/T\approx0.3$). Furthermore, many of these quenching events, regardless of timescale, lead to remnants with intermediate B/T, which do not fit cleanly into the two extreme scenarios mentioned previously. This is already a hint that the connection between transition timescales and final morphology is non-trivial, at least in the SAM. 

\begin{figure} 
\begin{center}
\includegraphics[width=\hsize]{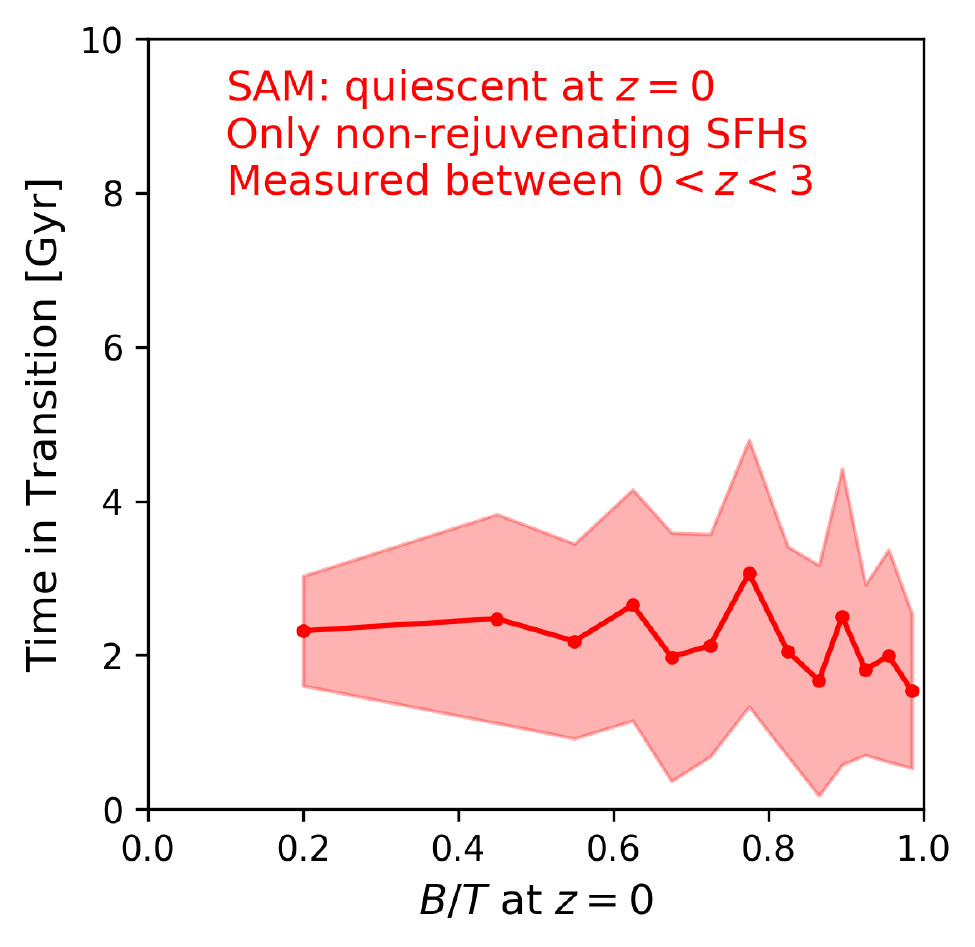}
\end{center}
\caption{The total time since $z=3$ that galaxies in the SAM (classified as quiescent at $z=0$) have spent in the transition region, as a function of their $z=0$ B/T ratio. We have restricted this analysis only to non-rejuvenating SFHs since rejuvenation events would artificially increase the B/T ratio and the total time spent in the transition region. The red circles and line show the mean value in each B/T bin, and the shading reflects the standard error on the mean in each bin. The flatness and scatter of the relation does not agree with the simple expectation that the most heavily bulge-dominated galaxies at $z=0$ should have preferentially spent the least amount of time in the transition region since $z=3$.}
\label{fig:morphtime}
\end{figure}

In \autoref{fig:morphtime}, we show the fraction of time since $z=3$ that galaxies spent in the transition region, as a function of their B/T ratio at $z=0$. We only focus on galaxies that are quiescent at $z=0$ because we know that they actually quenched. We further restrict this analysis only to non-rejuvenating SFHs because we want a clean estimate of the quenching timescale and final morphology, whereas rejuvenation events will preferentially increase both the B/T ratio and the total time spent in the transition region. \autoref{fig:morphtime} does not reveal the negative correlation expected from the two simple scenarios depicted above; namely, that galaxies with the highest B/T values at $z=0$ should have spent the least amount of time in the transition region. Instead, there is significant scatter in the transition timescale for each B/T bin, and the average values are consistent with being flat. We surprisingly find a similar trend for the $z=0$ transition population, but that population is harder to interpret because it has not yet quenched and some fraction of it could arise from SFMS oscillations.

How do we reconcile the above with our intuitive expectation that stellar disks can slowly fade and redden without undergoing significant bulge growth? In the SAM, effectively all quenching events (even very slow ones; e.g., T1478 in \autoref{fig:repsfh}) are triggered by a merger, and there is thus some degree of bulge growth, regardless of how small (disk instabilities also play a prominent role for bulge growth, but mostly for galaxies on the SFMS and moreso at early times). Broadly considered, quiescent disk-dominated galaxies in our SAM are not the quiescent analogs of effectively pure-disk star-forming galaxies, as the former do harbor some relic bulge component, no matter how sub-dominant \citep[see also the discussion in][]{brennan15}. One of the reasons that the existence of ``faded" pure-disk quiescent galaxies in the real Universe would be surprising is that even a slowly-evolving, ``completely isolated" disk might be expected to undergo secular processes like bar formation and disk instabilities, which may build up a pseudo-bulge component, especially on cosmological timescales \citep[see the reviews by][]{kormendy04,fisher16}. 

In the important observational studies of \citet{bundy10} and \citet{masters1006}, it was noted that ``passive disk" galaxies still tend to harbor some degree of centrally concentrated light (i.e., they do not preferentially have $B/T\sim0$). \citet{lackner12} carried out astrophysically motivated bulge-disk decompositions on tens of thousands of galaxies at $z<0.05$, with the goal of studying the relative distribution of classical and pseudo-bulges among the blue cloud, green valley, and red sequence. Among many interesting results, they found that very few red sequence galaxies have $B/T\sim0$ (see their Figure 34), but that $\sim17\%$ of red sequence galaxies were consistent with hosting a pseudo-bulge (see their section 5.6). They also found that red sequence galaxies that were best fit with a bulge+disk model had significantly redder disk colors than green valley galaxies that were best fit with a bulge+disk model (these galaxies were not well fit by a pure exponential profile). While this does imply that at least some fraction of red sequence galaxies with intermediate B/T underwent ``disk fading" rather than bulge growth as part of their quenching process, it does not fully explain the origin of the ``pre-existing" bulge component in these composite bulge+disk systems.

Our exploratory analysis spanning $0<z<3$ is complementary to and builds on the seminal observational studies of \citet{schawinski14} and \citet{smethurst15}, which addressed the diversity of pathways through the classical green valley at $z\sim0.1$ while also taking into account morphology. As we showed in \autoref{fig:medianz}, transition galaxies in both the observations and the SAM (out to $z=3$) do not seem to be preferentially extremely disk-dominated or extremely bulge-dominated \citep[as implied by][]{schawinski14}; instead they tend to have intermediate S{\'e}rsic index values, which suggests that both the disk and bulge exhibit significant amounts of light \citep[note also that composite bulge+disk galaxies dominate the green valley at $z<0.05$ based on the work of][]{lackner12}. Interpreting the cosmological origin and evolution of these intermediate B/T systems has historically been a very difficult task \citep[e.g., see the classic review by][]{dressler84}. In the SAM, galaxies with intermediate B/T (at $z=0$) have diverse evolutionary histories: they undergo quenching, rejuvenation, and morphological change on a variety of timescales, and their bulges can be built up through both mergers and disk instabilities. This diversity is qualitatively in agreement with the results of \citet{smethurst15}, who found that their observational sample of $z\sim0.1$ galaxies with intermediate B/T was consistent with a continuum of quenching timescales (conditional on their universal assumption of exponentially-declining single-$\tau$ SFHs). 

\subsection{Predictions for Non-structural Properties}
\begin{figure*} 
\begin{center}
\includegraphics[width=\hsize]{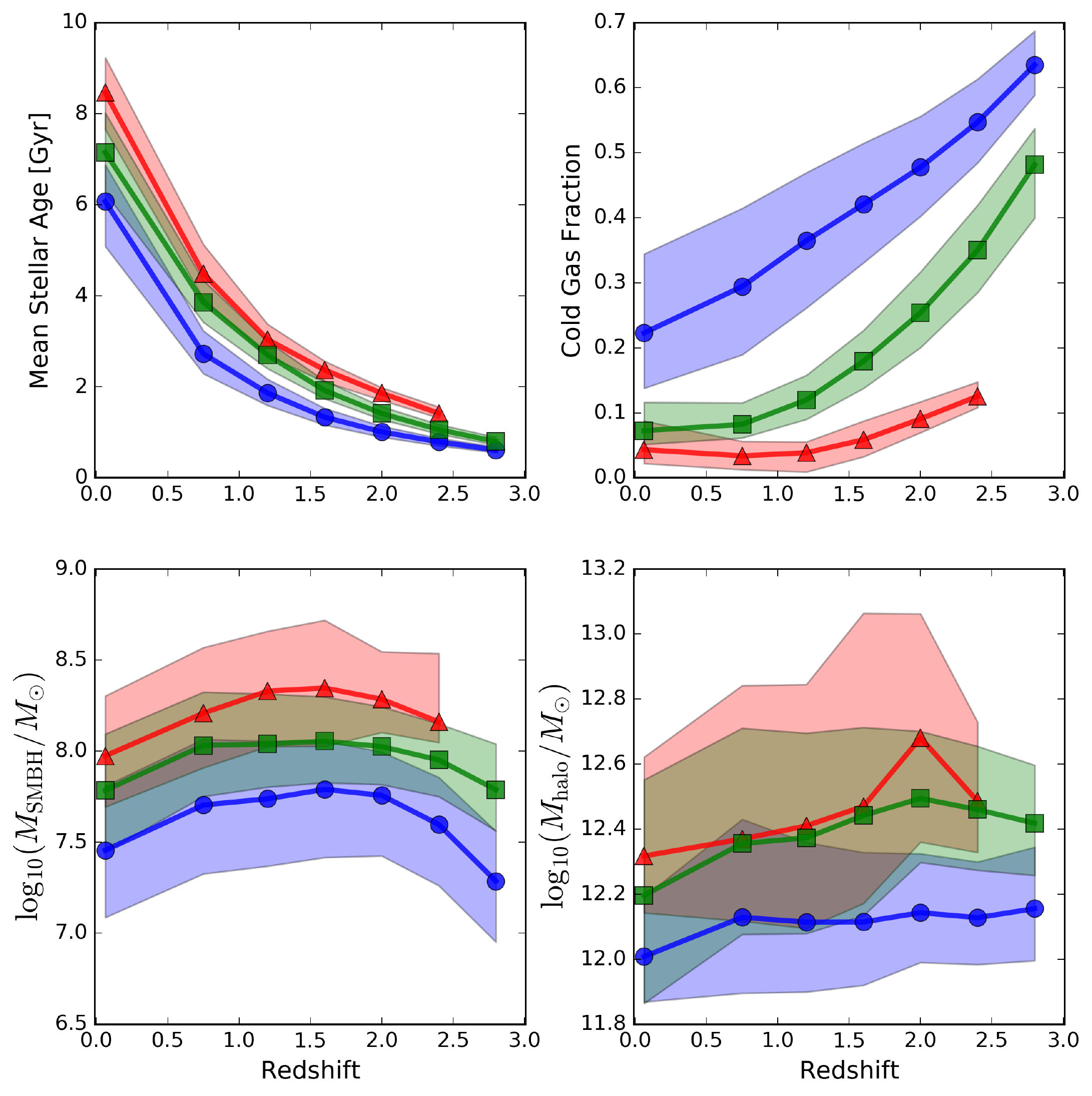}
\end{center}
\caption{Predictions from the SAM for the redshift evolution of additional non-structural properties for star-forming (blue), transition (green) and quiescent (red) galaxies: mean stellar age (top-left), cold gas fraction (top-right), SMBH mass (bottom-left), and dark matter halo mass (bottom-right). These predictions are from the cosmologically representative SAM light cones (described in \autoref{sec:samdesc}), are restricted to central galaxies only, and are based on the stellar mass matching algorithm described in \autoref{sec:mmatch}. The quiescent predictions have been truncated at $z=2.6$ due to the low number of quiescent galaxies in the highest SAM redshift slice. Note the striking separation between the predictions for the three subpopulations, and the preference that transition galaxies have for intermediate values of these non-structural properties.}
\label{fig:sampred}
\end{figure*}

Throughout this paper, we have mostly focused on comparing the structural properties of galaxies in the observations and the SAM. Although the SAM does not quantitatively reproduce the observations in several respects (e.g., the quiescent fraction at high redshift), it is still worthwhile to make some predictions so that future observations can try to test the general paradigm of the transition population. In \autoref{fig:sampred}, we present predictions from the SAM for the redshift evolution of four non-structural properties: the mean stellar age, the cold gas fraction, the SMBH mass, and the halo mass. Note that we have controlled for stellar mass dependence when comparing star-forming, transition and quiescent galaxies (just like in \autoref{fig:medianz}). 

The SAM predicts the mass-weighted mean stellar ages of transition galaxies to be intermediate between those of star-forming and quiescent galaxies (top-left panel of \autoref{fig:sampred}). Deep rest-frame UV-optical spectroscopy could be used to derive non-parametric SFHs, and thus mean stellar ages of mass-matched samples of star-forming, transition and quiescent galaxies with the goal of establishing a dominant evolutionary sequence. In such an observational evolutionary sequence, transition galaxies should have older stellar populations than star-forming galaxies but younger stellar populations than quiescent galaxies. Placing robust constraints on the mean stellar age is tremendously difficult because of the dust-age-metallicity degeneracy but it is an interesting target for future infrared and spectroscopic observing campaigns \citep[see also, e.g.,][]{whitaker10,gallazzi14,fumagalli16}.

More directly related to the intermediate suppression of star formation in transition galaxies (relative to what we call quiescent galaxies) is the cold gas fraction ($\equiv\frac{M_{\rm cold}}{M_{\rm cold}+M_*}$). Not surprisingly, our models predict that the cold gas fractions of transition galaxies are intermediate between those of star-forming and quiescent galaxies (top-right panel of \autoref{fig:sampred}). An interesting question observationally is whether the star formation efficiency ($\equiv\frac{SFR}{M_{\rm cold}}$) in transition galaxies exhibits a similar trend, and whether the lower amount of cold gas in transition galaxies (relative to star-forming galaxies) is due to stronger feedback or lower gas accretion rates. Modern and future facilities such as the \textit{Atacama Large Millimeter Array} may be useful in linking the cold gas fractions and star formation efficiencies of transition galaxies to feedback events and other physical mechanisms responsible for quenching and morphological change \citep[see also][]{cortese09,alatalo14,french15,alatalo16,barro16b,spilker16}.

In our SAM, as in many cosmological hydrodynamic simulations, SMBHs play a prominent role in quenching massive galaxies. We predict that transition galaxies should host SMBHs that are intermediate in mass between those of star-forming and quiescent galaxies (bottom-left panel of \autoref{fig:sampred}). This suggests that SMBHs in transition galaxies are largely nearing the end of their growth (unlike the SMBHs that are still growing in star-forming galaxies and the SMBHs in quiescent galaxies that have minimal or no ongoing growth), thus making transition galaxies important observational targets for studying the shutdown of common SMBH growth channels \citep[see also, e.g.,][]{volonteri10,greene12,kocevski12,trump13,terrazas16,terrazas16b,azadi16}.

In this paper, we have not probed the role of the environment for producing transition and quiescent galaxies because our observations do not cover the dense regions where environmental effects are thought to dominate. Nevertheless, many studies have explored possible relationships between quenching and proxies for environment. One very relevant result from the literature is the tendency of classical green valley galaxies to live in intermediate density environments \citep[e.g.,][]{coil08,peng10,zehavi11,behroozi13,krause13,woo15,rodriguezpuebla15,mandelbaum16}. In our SAM, although transition galaxies tend to live in intermediate mass halos compared to stellar mass-matched samples of star-forming and quiescent galaxies out to $z\sim2.5$, there is significant overlap in the halo mass distributions of transition and quiescent galaxies (bottom-right panel of \autoref{fig:sampred}). This overlap may partially be explained by the fact that both transition and quiescent galaxies tend to be bulge-dominated, and that many of these bulges were built up through mergers, which lead to increased halo masses for the remnants. Intriguingly, the preference of transition galaxies to have slightly lower halo masses compared to quiescent galaxies is likely related to their preference for intermediate B/T ratios (\autoref{fig:medianz}) and greater likelihood of disk instability-driven bulge growth rather than merger-driven bulge growth \citep[since transition galaxies tend to have a more substantial disk component; see also][]{tonini16}. 

It is unlikely that all of our predictions for the redshift evolution of non-structural properties are exactly and quantitatively correct. However, these results still do qualitatively suggest that there are other non-morphological ways to probe the evolutionary significance of transition galaxies.

\section{Summary}\label{sec:summary}
We have carried out a comprehensive analysis of massive ``transition galaxies" with $M_*>10^{10}M_{\odot}$. These transition galaxies are defined in a physically and statistically motivated way to have intermediate sSFR values below the SFMS. Our investigation has been done on observations from the GAMA survey at $z\sim0.1$ and the CANDELS survey at $0.5<z<3.0$, as well as on a cosmologically representative semi-analytic model of galaxy formation and a hydrodynamical simulation with state-of-the-art mechanical AGN feedback. The main results of our paper are as follows:

\begin{enumerate}
\item In both the observations and the SAM, transition galaxies tend to have intermediate structural properties compared to star-forming and quiescent galaxies (after controlling for stellar mass). The three structural properties that we probe in this paper are the S{\'e}rsic index, the half-light radius, and the surface stellar mass density. One possible interpretation is that morphological change accompanies or precedes quenching because transition galaxies are not yet fully quenched, but they are already substantially more compact and concentrated than star-forming galaxies (and less so than quiescent galaxies). However, the ``progenitor bias" concept likely plays a non-trivial role: transition and quiescent galaxies might be more compact than star-forming galaxies in the same epoch simply because they began to quench at earlier times, when all galaxies were smaller (and because star-forming galaxies continue to grow more rapidly in size and mass).
\item The fraction of all galaxies that are in the transition region remains constant at $\sim20\%$ in the observations at $0.5<z<3.0$. In the SAM, this is also the case and is due to the fact that galaxies are constantly moving into and out of the transition region on a variety of timescales and from various directions. The SAM has a deficit of quiescent galaxies at $z>0.5$, but matches the observations very well at $z\sim0.1$. This suggests that the timescales on which galaxies enter and move through the transition region (i.e., quenching timescales) are too long in the SAM. 
\item We explicitly use the transition population that we identified in the observations to place an observational upper limit on the average population transition timescale as a function of redshift. This average transition timescale is consistent with ``fast track" quenching at high redshift ($\sim0.8$ Gyr at $z\sim2.5$), and ``slow track" quenching at low redshift ($\sim7$ Gyr at $z\sim0.5$). This is an upper limit because of systematic uncertainties in the observations and because we have made the extreme assumption that galaxies only transition once from the SFMS toward quiescence (i.e., without any rejuvenation). Our calculation can be refined in the future as more observational constraints become available.
\item We qualitatively identify four different evolutionary modes for the physical origin of transition galaxies in the SAM: oscillations on the SFMS, slow quenching, fast quenching, and rejuvenation. Each of these modes is driven by different or overlapping physical processes that act on different timescales, including mergers, disk instabilities, starbursts, and feedback from stars, supernovae and AGN.
\item The average $z=0$ quiescent galaxy in the SAM first began its quenching event at $z\sim0.7$ and spent an average of $\sim2$ Gyr in the transition region before quenching by $z\sim0.4$. Only $\sim20\%$ of $z=0$ quiescent galaxies in the SAM began their quenching event at $z>2$. The scarcity of high redshift quenching events along with the quenching timescales typically being too slow explains the deficit of high redshift quiescent galaxies in the SAM.
\item Cosmological hydrodynamical ``zoom in" simulations by \citet{choi16}, with state-of-the-art implementation of mechanical AGN feedback, are able to reproduce the truly ``red and dead" giant ellipticals that we observe at $z\sim0$. We find that these simulated galaxies (which span $M_{\rm halo}\sim10^{12-13.4}M_{\odot}$) tend to become quiescent by $z\sim2$, which suggests that a more sophisticated treatment of momentum-driven AGN feedback can help boost the quiescent fraction at high redshift in the SAM. However, even the fastest quenching galaxies in these hydrodynamical simulations do not reproduce the observed quiescent population at $z>2$, which is also a fundamental problem in our SAM and other simulations. Future studies will need to address how AGN feedback might be coupled to other quenching mechanisms at these early epochs to reproduce the heavily quiescent galaxies that we observe at $z\sim3$ (when the Universe is only $\sim2.2$ Gyr old).
\item We find the surprising result that, in the SAM, the time spent in the transition region since $z=3$ is independent of the $B/T$ ratio at $z=0$ for galaxies that are quiescent at $z=0$ and that have not experienced any rejuvenation. This is different from the negative correlation that we might expect between the quenching timescale and the $B/T$ ratio (i.e., the expectation that bulge-dominated quiescent galaxies quenched more quickly than disk-dominated quiescent galaxies). 
\item We use the SAM to predict the redshift evolution of the mean stellar ages, cold gas fractions, SMBH masses, and halo masses of star-forming, transition and quiescent galaxies since $z=3$ (massive central galaxies only). Transition galaxies tend to exhibit intermediate values of these properties relative to the star-forming and quiescent subpopulations (after controlling for stellar mass dependence). We therefore predict that these non-structural properties might offer additional ways to observationally test the general paradigm of the transition population. 
\end{enumerate}

In this paper, we have raised several important observational and theoretical questions about how galaxies might move below the SFMS at $0<z<3$. In the future, it will be important to test how different models (both semi-analytic and hydrodynamic) can be made to reproduce the full observed spread in the ``degree of quiescence" of galaxies below the SFMS as a function of redshift. This is important because it might reveal clues about the timescales on which galaxies quench and rejuvenate, and the relative frequency of such transitions. 

\section*{Acknowledgements}
We thank the anonymous referee for suggestions that improved the manuscript. We further thank James E. Gunn, Barry Madore, Jennifer Lotz, Romeel Dave, Phil Hopkins, Lars Hernquist, Aaron Romanowsky, Irene Shivaei, Sandro Tacchella and Samir Salim for inspiring discussions. VP is grateful for the encouragement and support from Roderich Tumulka, the astronomy faculty at Rutgers (in particular Saurabh Jha, Andrew Baker and Tad Pryor), the Princeton Post-Baccalaureate Program in Astrophysics (in particular Jenny Greene, Anatoly Spitkovsky, Michael Strauss, Jill Knapp and Jesus Hinojosa), and John Mulchaey. RB was supported in part by HST Theory grant HST-AR-13270-A. RSS thanks the Downsbrough family for their generous support, and acknowledges support from the Simons Foundation through a Simons Investigator grant. PB was supported by program number HST-HF2-51353.001-A, provided by NASA through a Hubble Fellowship grant from the Space Telescope Science Institute, which is operated by the Association of Universities for Research in Astronomy, Incorporated, under NASA contract NAS5-26555. AK gratefully acknowledges support from the YCAA Prize Postdoctoral Fellowship. We acknowledge the contributions of hundreds of individuals to the planning and support of the CANDELS observations, and to the development and installation of new instruments on HST, without which this work would not have been possible. Support for Program number HST-GO-12060 was provided by NASA through a grant from the Space Telescope Science Institute, which is operated by the Association of Universities for Research in Astronomy, Incorporated, under NASA contract NAS5-26555. This work has made use of the Rainbow Cosmological Surveys Database, which is operated by the Universidad Complutense de Madrid (UCM), partnered with the University of California Observatories at Santa Cruz (UCO/ Lick, UCSC).

\bibliographystyle{mnras}
\bibliography{references}


\appendix
\section{Optical Attenuation in the $UVJ$ Diagram}\label{sec:uvjav}
Here we decompose the $UVJ$ diagram for each of our CANDELS redshift slices based on our sSFR-$M_*$ subpopulation classification (star-forming, transition and quiescent). We then color-code the points by their best-fit optical attenuation $A_V$, which is output from SED fitting as described in \autoref{sec:datadesc}. Clearly, star-forming galaxies occupy a ``dust sequence" \citep[e.g.,][]{wuyts07,williams09,brammer11}, and the quiescent galaxies tend to remain within the empirical ``quiescent wedge" \citep[boundary equations taken from][]{vandokkum15}. A non-negligible number of our sSFR-M$_*$-defined transition galaxies extend into the classical dusty star-forming region, with high rest-frame $(U-V)$ and $(V-J)$ colors, even though their $A_V$ values typically do not approach the large values found for classical dusty star-forming galaxies. Future work will be needed to determine whether this is indeed a population of ``dusty transition galaxies," and what the implications of this population are for: (1) the full diversity of transition pathways, including dusty post-starburst systems, and (2) systematic uncertainties in SED-based dust correction methods.

\begin{figure*} 
\begin{center}
\includegraphics[width=\hsize]{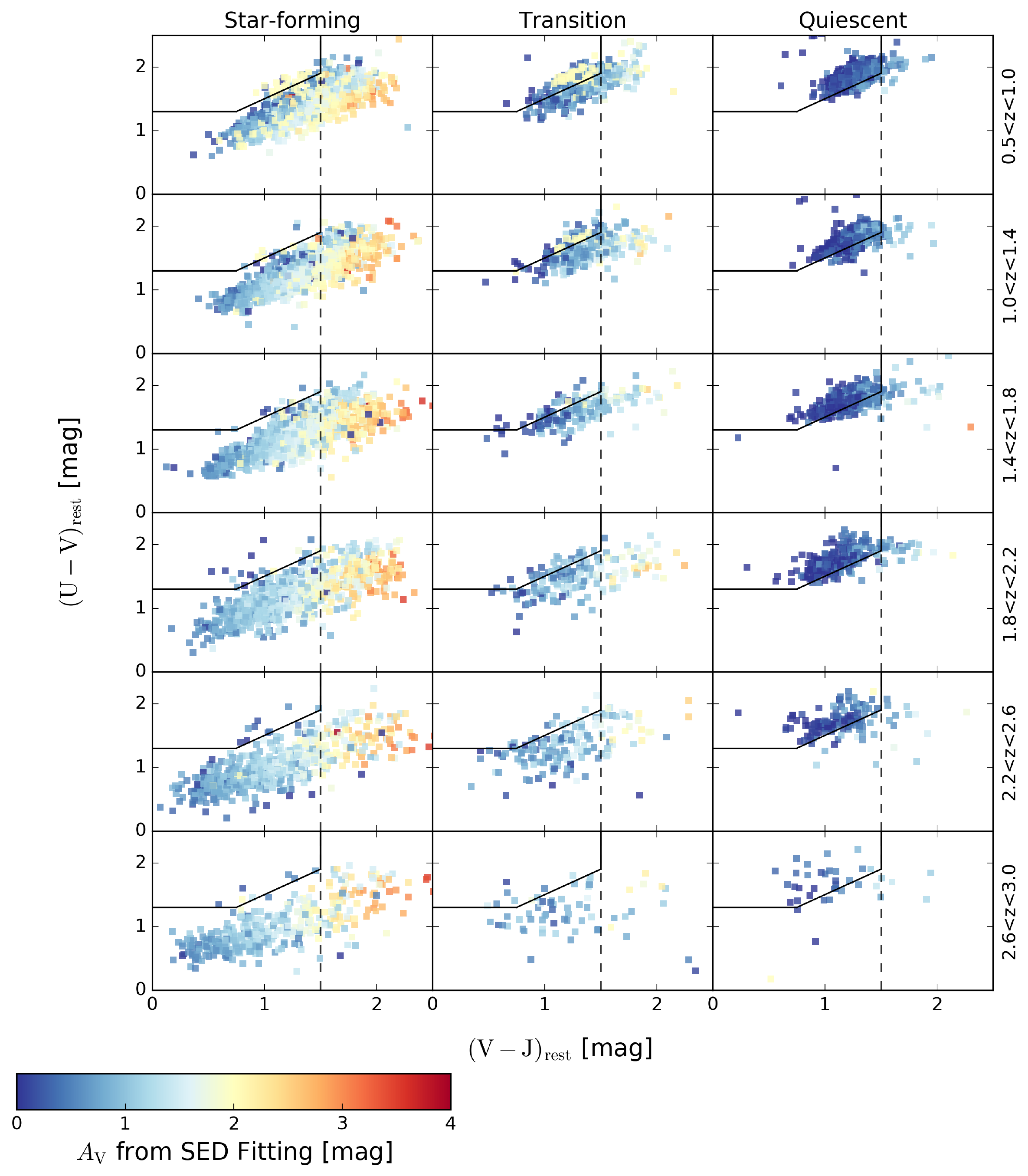}
\end{center}
\caption{For completeness, here we show the $UVJ$ distribution of sSFR-M$_*$-selected star-forming, transition and quiescent galaxies in each CANDELS redshift slice using different subpanels. We also color-code the points by the best-fit $A_V$ that is output by SED fitting. The boundaries for the empirical ``quiescent wedge" are taken directly from \citet{vandokkum15}. Note how the star-forming galaxies form a ``dust sequence" and tend to stay outside of the quiescent wedge, whereas the quiescent galaxies tend to stay within the empirically-defined quiescent wedge and are relatively dust-free \citep[e.g.,][]{wuyts07,williams09,brammer11}. In contrast, our sSFR-M$_*$-selected transition galaxies tends to span the region between the quiescent and star-forming galaxies, and a non-negligible number of transition galaxies extend into the classical dusty star-forming region, with high rest-frame $(U-V)$ and $(V-J)$ colors (although their $A_V$ do not typically approach the large values of classical dusty star-forming galaxies).}
\label{fig:uvjav}
\end{figure*}

\section{Cumulative Distribution Functions of Structural Properties}\label{sec:cdf}
\begin{figure*} 
\begin{center}
\includegraphics[width=0.6\hsize]{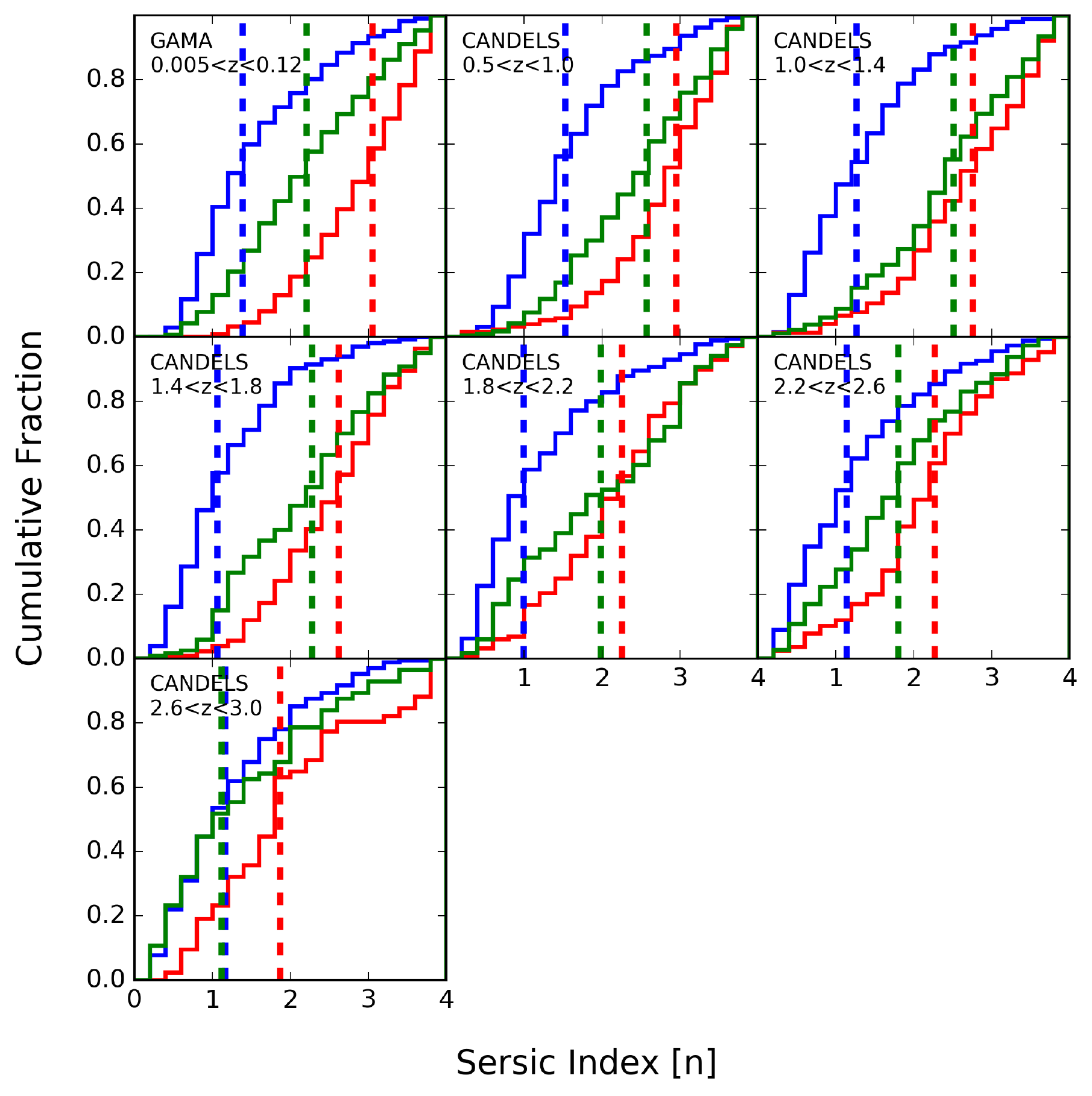}
\includegraphics[width=0.6\hsize]{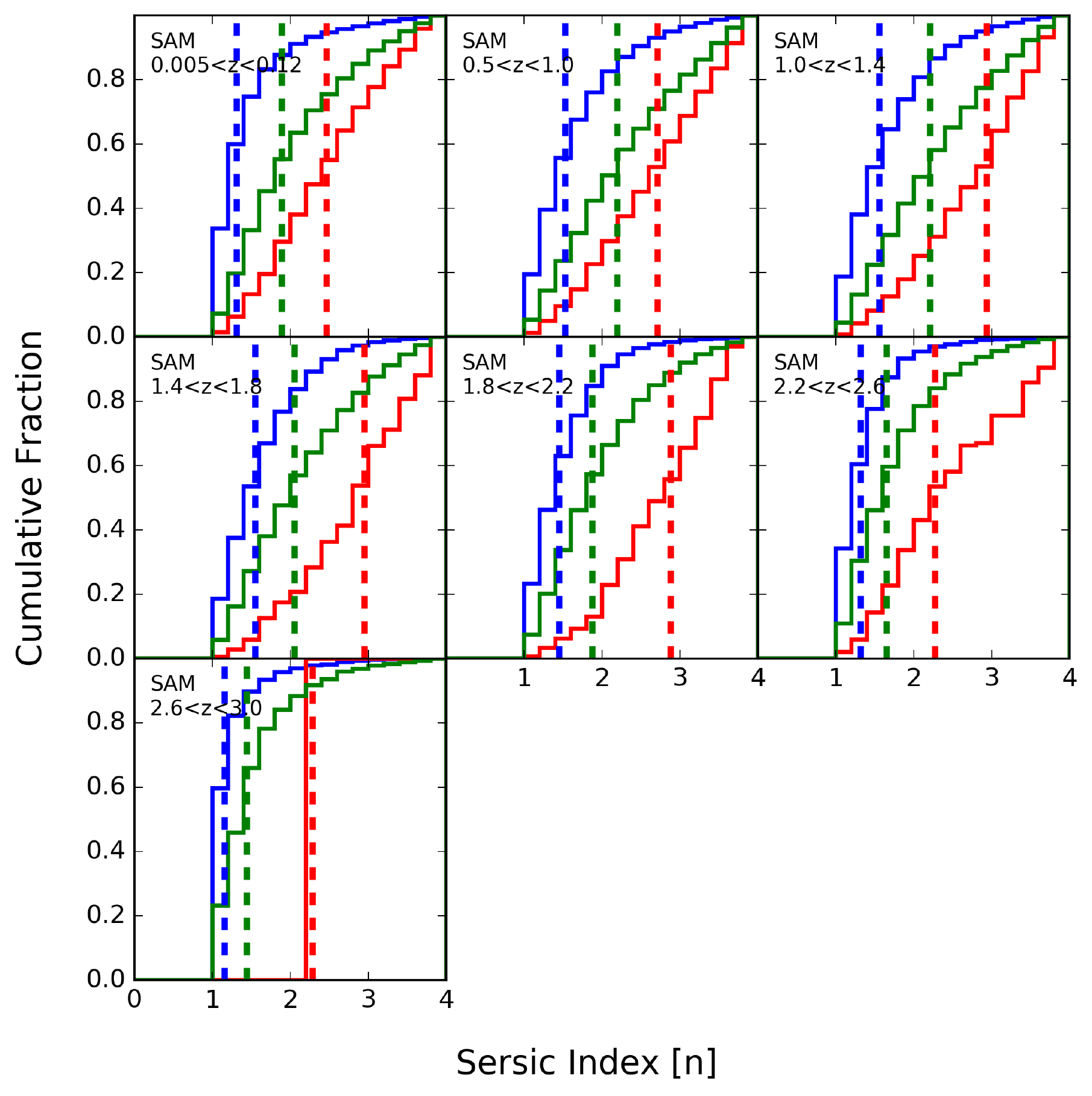}
\end{center}
\caption{Normalized cumulative distribution functions for the S{\'e}rsic index of galaxies in the observations (top) and the SAM (bottom) in each of our redshift slices. The CDFs have been split into three separate ones for galaxies that are classified as star-forming (blue), transition (green), and quiescent (red). The dashed vertical lines mark the median value of each CDF in every panel.}
\label{fig:cdfsersic}
\end{figure*}

\begin{figure*} 
\begin{center}
\includegraphics[width=0.6\hsize]{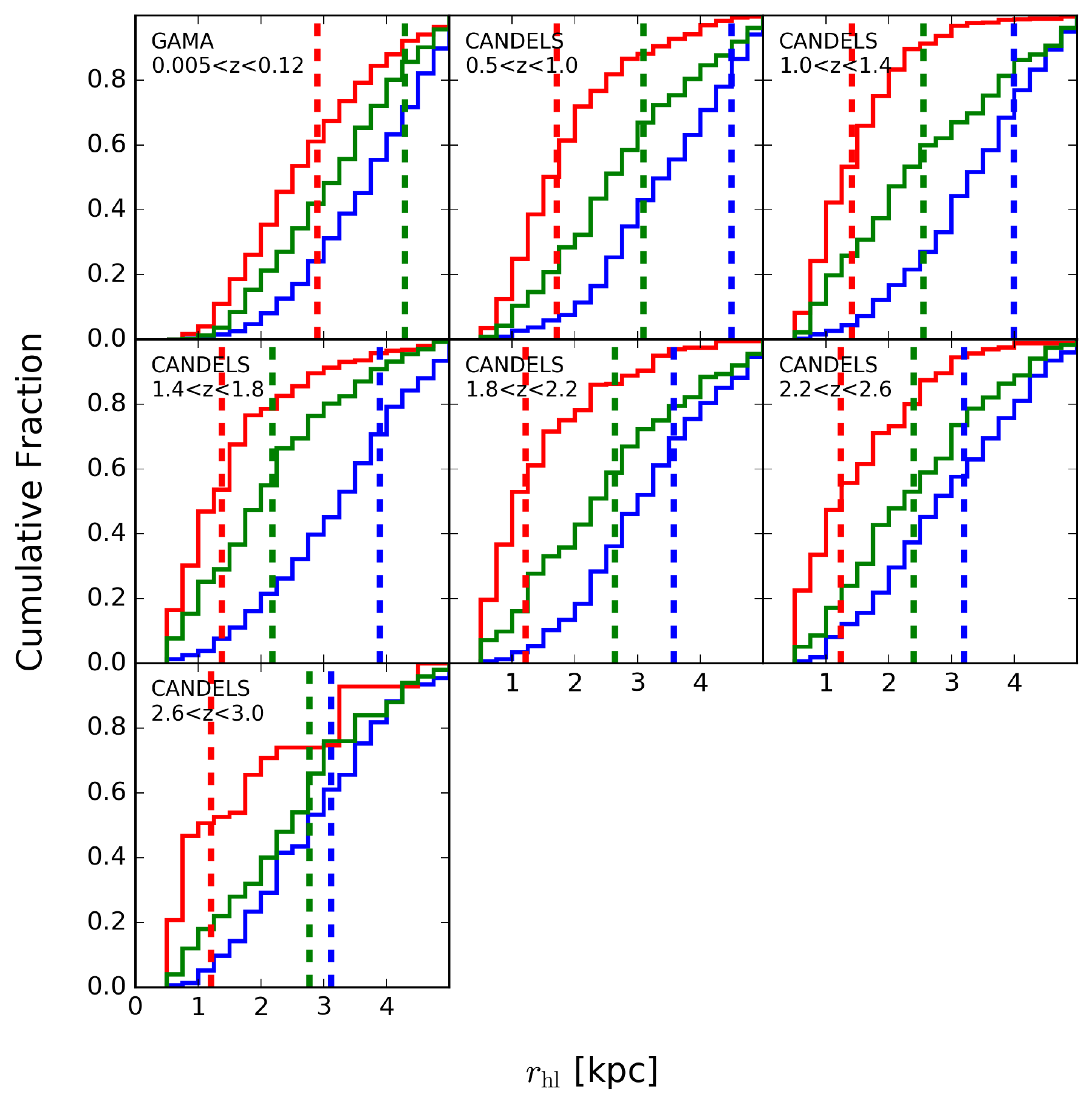}
\includegraphics[width=0.6\hsize]{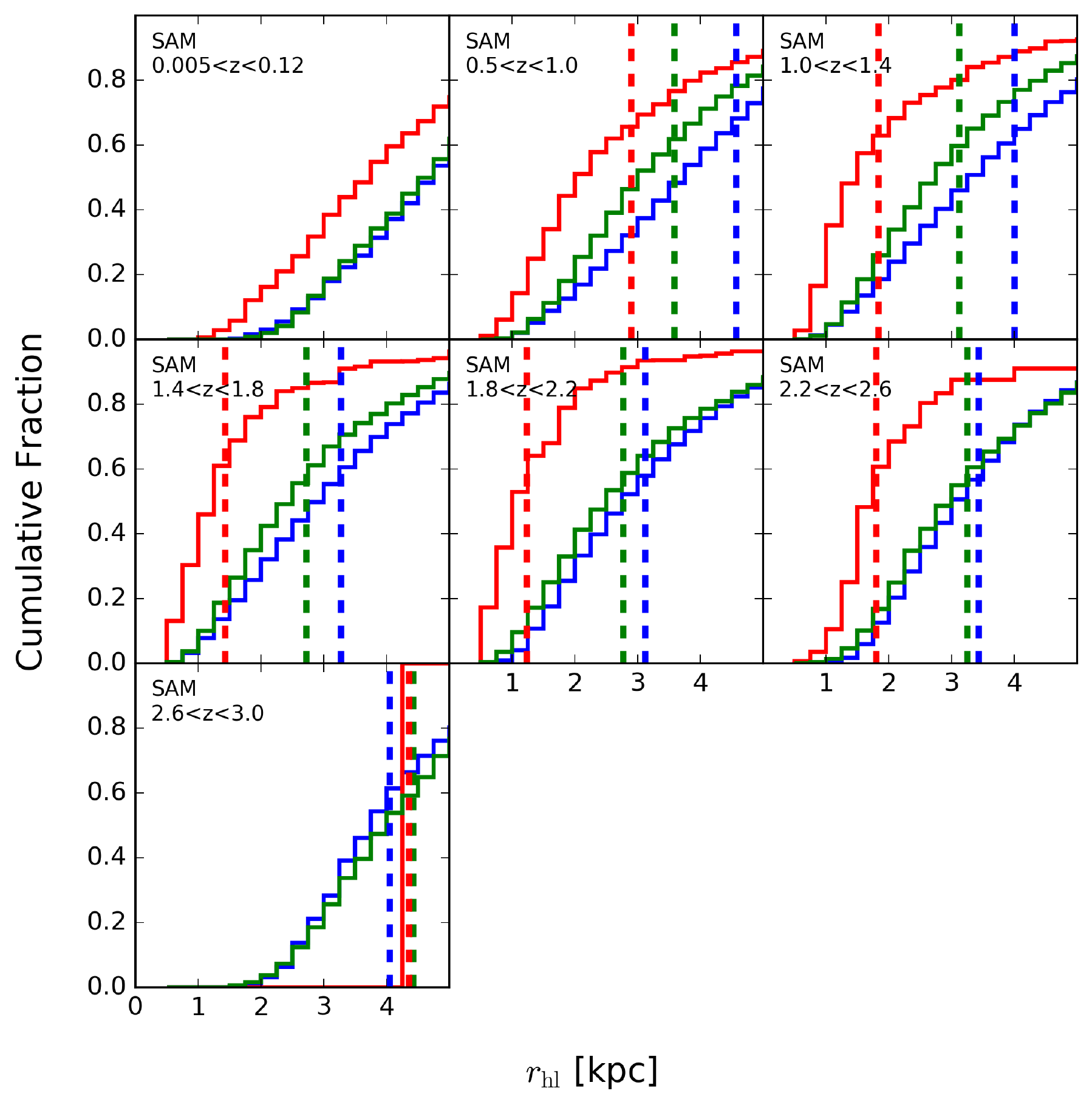}
\end{center}
\caption{Normalized cumulative distribution functions for the half-light radius of galaxies in the observations (top) and the SAM (bottom) in each of our redshift slices. The CDFs have been split into three separate ones for galaxies that are classified as star-forming (blue), transition (green), and quiescent (red). The dashed vertical lines mark the median value of each CDF in every panel.}
\label{fig:cdfrhl}
\end{figure*}

\begin{figure*} 
\begin{center}
\includegraphics[width=0.6\hsize]{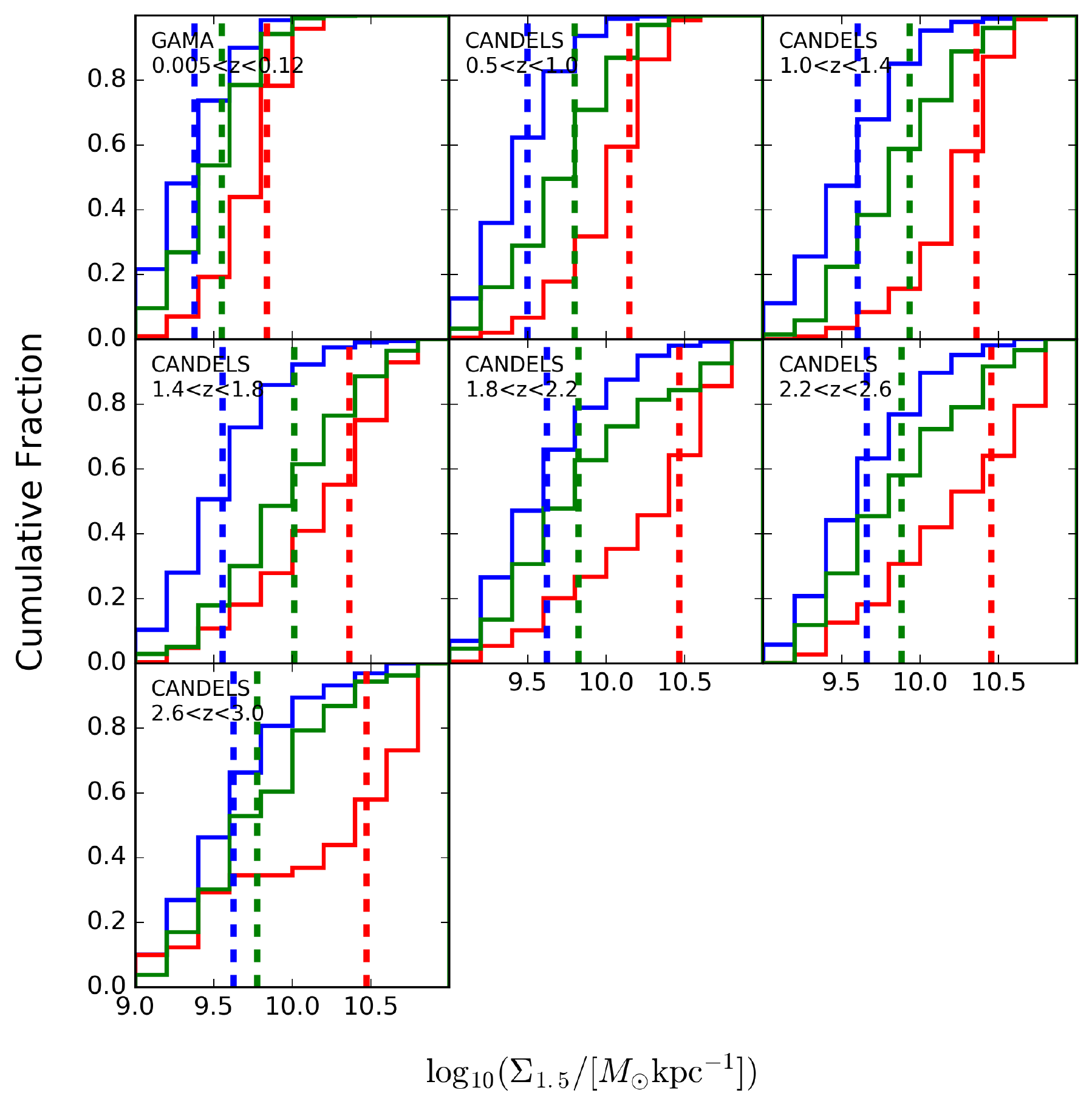}
\includegraphics[width=0.6\hsize]{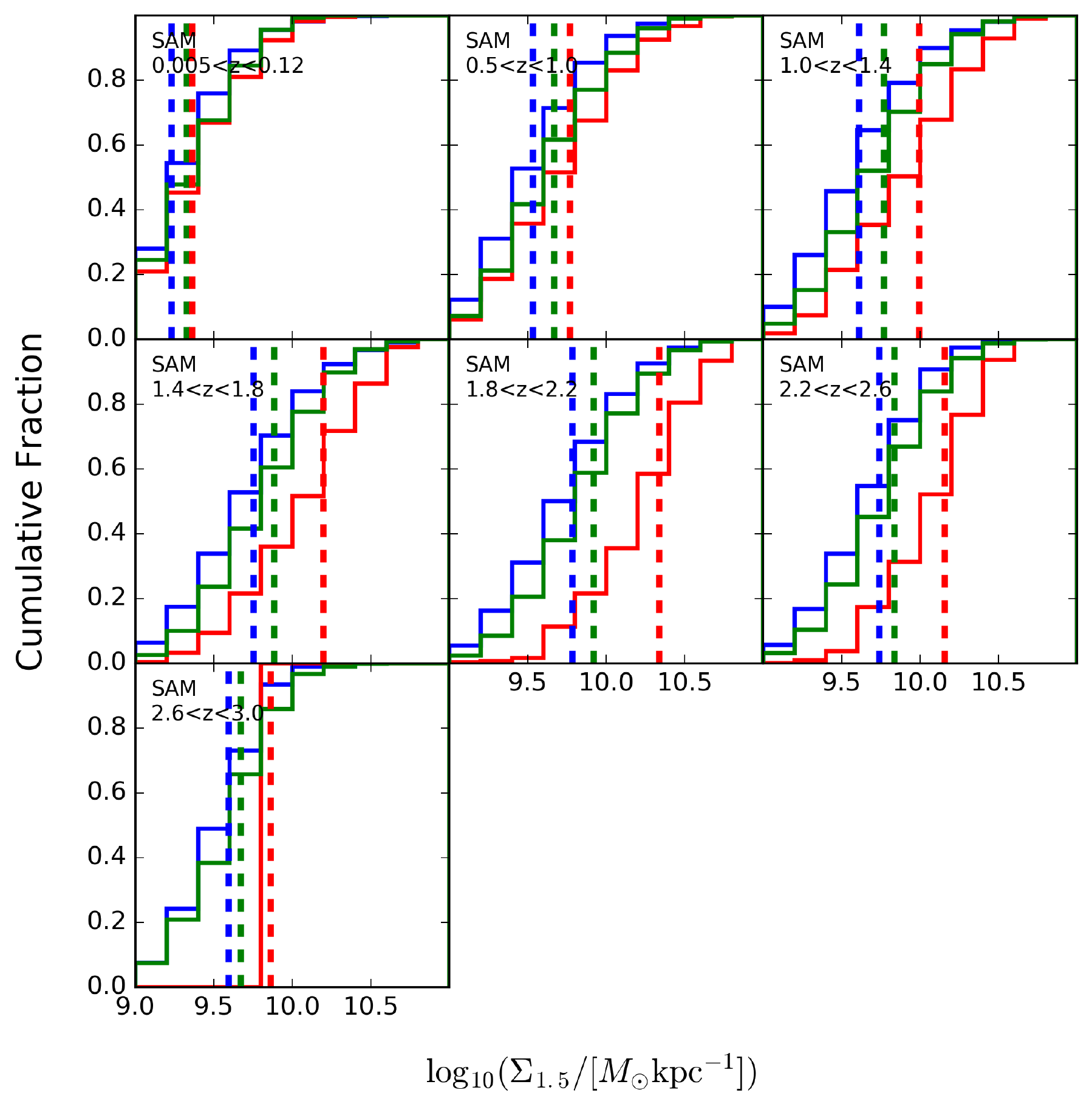}
\end{center}
\caption{Normalized cumulative distribution functions for the $\Sigma_{1.5}$ parameter of galaxies in the observations (top) and the SAM (bottom) in each of our redshift slices. The CDFs have been split into three separate ones for galaxies that are classified as star-forming (blue), transition (green), and quiescent (red). The dashed vertical lines mark the median value of each CDF in every panel.}
\label{fig:cdfs15}
\end{figure*}

Here we show the full cumulative distribution functions for the structural properties of galaxies in the observations and the SAM. The results presented in \autoref{fig:medianz} for the redshift evolution of the S{\'e}rsic index, half-light radius, and $\Sigma_{1.5}$ are based on the cumulative distribution functions shown in \autoref{fig:cdfsersic}, \autoref{fig:cdfrhl}, and \autoref{fig:cdfs15}, respectively. 

We also ran two-sample Kolmogorov-Smirnov tests to compare transition galaxies' structural properties to those of mass-matched star-forming and quiescent galaxies. The resulting $p$-values are $\ll0.001$ in a majority of cases, as shown in \autoref{tab:ksobs} and \autoref{tab:kssam}. This suggests that transition galaxies' structural property distributions across a wide redshift range are drawn from different parent populations compared to those of mass-matched star-forming and quiescent galaxies. 

\begin{table*}
\begin{tabular}{cccccccc}\hline
Redshift Slice & $n^{\rm T,SF}$ & $n^{\rm T,Q}$ & $r_{\rm hl}^{\rm T,SF}$ & $r_{\rm hl}^{\rm T,Q}$ & $\Sigma_{1.5}^{\rm T,SF}$ & $\Sigma_{1.5}^{\rm T,Q}$ & Sample \\
(1) & (2) & (3) & (4) & (5) & (6) & (7) & (8) \\\hline
0.005<z<0.12 & $\ll0.001$ & $\ll0.001$ & $\ll0.001$ & $\ll0.001$ & $\ll0.001$ & $\ll0.001$ & GAMA \\
0.5<z<1.0 & $\ll0.001$ & $\ll0.001$ & $\ll0.001$ & $\ll0.001$ & $\ll0.001$ & $\ll0.001$ & CANDELS \\
1.0<z<1.4 & $\ll0.001$ & 0.008 & $\ll0.001$ & $\ll0.001$ & $\ll0.001$ & $\ll0.001$ & CANDELS \\
1.4<z<1.8 & $\ll0.001$ & $\ll0.001$ & $\ll0.001$ & $\ll0.001$ & $\ll0.001$ & $\ll0.001$ & CANDELS \\
1.8<z<2.2 & $\ll0.001$ & 0.004 & $\ll0.001$ & $\ll0.001$ & $\ll0.001$ & $\ll0.001$ & CANDELS \\
2.2<z<2.6 & $\ll0.001$ & $\ll0.001$ & $\ll0.001$ & $\ll0.001$ & $\ll0.001$ & $\ll0.001$ & CANDELS \\
2.6<z<3.0 & 0.564 & $\ll0.001$ & 0.191 & $\ll0.001$ & 0.041 & $\ll0.001$ & CANDELS \\
\end{tabular}
\caption{Two-sample Kolmogorov-Smirnov test $p$-values for structural distinctiveness of transition galaxies in the observations. The names of columns 2-7 indicate the structural property (S{\'e}rsic index, half-light radius or mass surface pseudodensity) and which mass-matched subpopulations are being compared (transition versus star-forming, or transition versus quiescent galaxies). The values in the table are the $p$-values from the two-sample Kolmogorov-Smirnov test. If a $p$-value is less than $10^{-3}$, we show $\ll0.001$ instead.}
\label{tab:ksobs}
\end{table*}

\begin{table*}
\begin{tabular}{cccccccc}\hline
Redshift Slice & $n^{\rm T,SF}$ & $n^{\rm T,Q}$ & $r_{\rm hl}^{\rm T,SF}$ & $r_{\rm hl}^{\rm T,Q}$ & $\Sigma_{1.5}^{\rm T,SF}$ & $\Sigma_{1.5}^{\rm T,Q}$ & Sample \\
(1) & (2) & (3) & (4) & (5) & (6) & (7) & (8) \\\hline
0.005<z<0.12 & $\ll0.001$ & $\ll0.001$ & $\ll0.001$ & $\ll0.001$ & $\ll0.001$ & 0.025 & SAM \\
0.5<z<1.0 & $\ll0.001$ & $\ll0.001$ & $\ll0.001$ & $\ll0.001$ & $\ll0.001$ & $\ll0.001$ & SAM \\
1.0<z<1.4 & $\ll0.001$ & $\ll0.001$ & $\ll0.001$ & $\ll0.001$ & $\ll0.001$ & $\ll0.001$ & SAM \\
1.4<z<1.8 & $\ll0.001$ & $\ll0.001$ & $\ll0.001$ & $\ll0.001$ & $\ll0.001$ & $\ll0.001$ & SAM \\
1.8<z<2.2 & $\ll0.001$ & $\ll0.001$ & $\ll0.001$ & $\ll0.001$ & $\ll0.001$ & $\ll0.001$ & SAM \\
2.2<z<2.6 & $\ll0.001$ & $\ll0.001$ & $\ll0.001$ & $\ll0.001$ & $\ll0.001$ & $\ll0.001$ & SAM \\
2.6<z<3.0 & $\ll0.001$ & $-$ & $\ll0.001$ & $-$ & $\ll0.001$ & $-$ & SAM \\
\end{tabular}
\caption{Same as \autoref{tab:ksobs} but for the SAM instead. Since there are so few quiescent galaxies in the SAM at $2.6<z<3.0$, we do not run the two-sample Kolmogorov-Smirnov tests comparing transition to quiescent galaxies' distributions in that redshift slice.}
\label{tab:kssam}
\end{table*}

\section{Observed Galaxy Number Densities}\label{sec:numdensity}
\begin{figure*} 
\begin{center}
\includegraphics[width=\hsize]{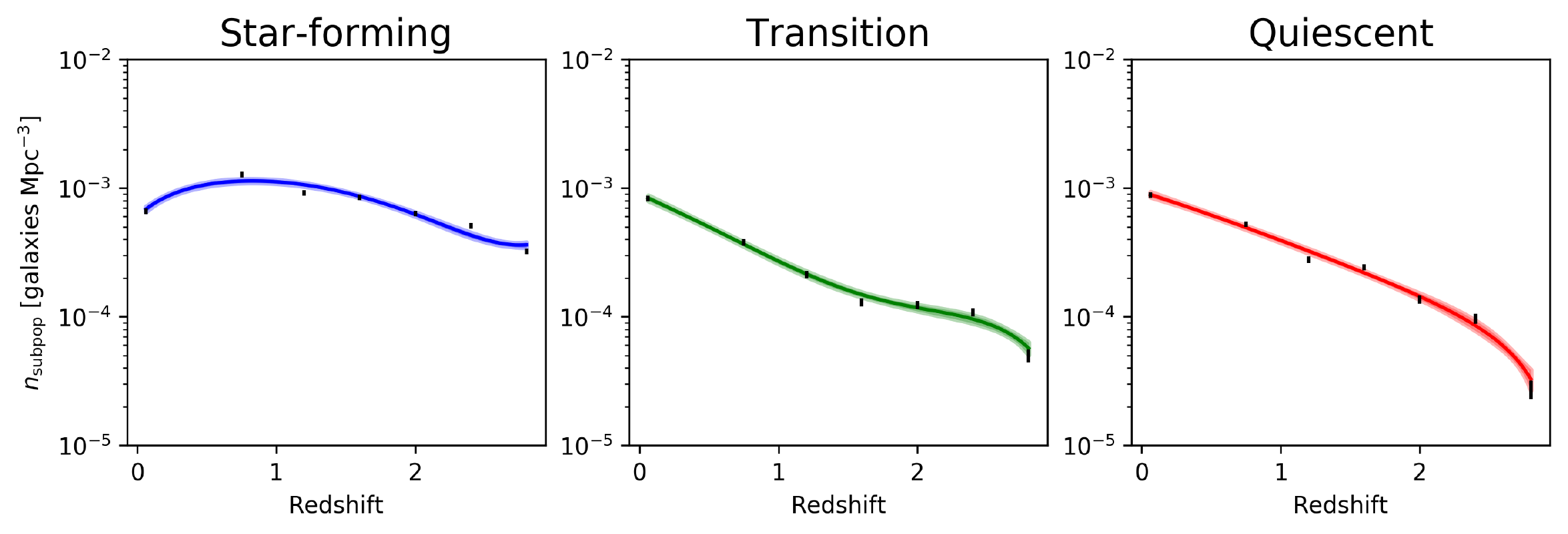}
\end{center}
\caption{The observed number densities of star-forming (left), transition (middle), and quiescent (right) galaxies as a function of redshift. The scatter points show our measurements with bootstrapped errorbars, the solid lines are our median cubic polynomial fits out of 1000 random trials, and the shading reflects the $16-84$ percentile spread in the random polynomial fits. The transition and quiescent galaxy polynomial fits are used to compute an observational upper limit on the average population transition timescale as a function of redshift in \autoref{sec:transit}.}
\label{fig:numdensity}
\end{figure*}

In \autoref{fig:numdensity}, we show the observed number densities of star-forming, transition and quiescent galaxies as a function of redshift (the actual values are given in \autoref{tab:transit}). We also plot our cubic polynomial fits to these number densities: 1000 random fits were done, and the median and $16-84$ percentile values as a function of redshift were recorded. The smooth cubic polynomial fits are used to calculate the average population transition timescale as a function of redshift using \autoref{eq:transit} in \autoref{sec:transit}. 

\begin{table}
\caption{Observed number densities of star-forming, transition and quiescent galaxies as a function of redshift. The uncertainties were obtained from bootstrapping of the sSFR-M$_*$ diagram, with a minimum systematic fractional uncertainty of five percent \citep[e.g.,][]{muzzin13}.}
\centering
  \begin{tabular}{cccc}
  \hline
Redshift & Star-forming & Transition & Quiescent \\
 & $10^{-4}$ Mpc$^{-3}$ & $10^{-4}$ Mpc$^{-3}$ & $10^{-4}$ Mpc$^{-3}$ \\\hline
0.005$<z<$0.12 & 6.57$\pm$0.33 & 8.31$\pm$0.42 & 8.83$\pm$0.44 \\
0.5$<z<$1.0 & 12.76$\pm$0.64 & 3.78$\pm$0.22 & 5.22$\pm$0.26 \\
1.0$<z<$1.4 & 9.15$\pm$0.46 & 2.13$\pm$0.14 & 2.76$\pm$0.17 \\
1.4$<z<$1.8 & 8.50$\pm$0.43 & 1.30$\pm$0.10 & 2.43$\pm$0.14 \\
1.8$<z<$2.2 & 6.33$\pm$0.32 & 1.23$\pm$0.09 & 1.36$\pm$0.10 \\
2.2$<z<$2.6 & 5.07$\pm$0.25 & 1.08$\pm$0.08 & 0.96$\pm$0.08 \\
2.6$<z<$3.0 & 3.21$\pm$0.16 & 0.50$\pm$0.06 & 0.27$\pm$0.05 \\
  \end{tabular}
  \label{tab:transit}
\end{table}

\section*{Affiliations}
$^1$Department of Astrophysical Sciences, Peyton Hall, Princeton University, Princeton, NJ 08544, USA\\
$^2$Department of Physics and Astronomy, Rutgers, The State University of New Jersey, 136 Frelinghuysen Rd, Piscataway, NJ 08854, USA\\
$^3$UCO/Lick Observatory, Department of Astronomy and Astrophysics, University of California, Santa Cruz, CA 95064, USA; \href{mailto:viraj.pandya@ucsc.edu}{viraj.pandya@ucsc.edu}\\
$^4$Department of Astronomy, University of California, Berkeley, CA 94720, USA\\
$^5$Department of Physics, University of Bath, Claverton Down, Bath, BA2 7AY, UK\\
$^6$Centre for Astrophysics and Supercomputing, Swinburne University of Technology, PO Box 218, Hawthorn 3122, Australia\\
$^7$Hubble Fellow\\
$^8$Yale Center for Astronomy \& Astrophysics, Physics Department, P.O. Box 208120, New Haven, CT 06520, USA\\
$^{9}$Department of Physics, University of California at Santa Cruz, Santa Cruz, CA 95064, USA\\
$^{10}$Department of Physics and Astronomy, University of Missouri-Kansas City, 5110 Rockhill Road, Kansas City, MO 64110, USA\\
$^{11}$Department of Physics and Astronomy, Colby College, Waterville, ME 04901, USA\\
$^{12}$Department of Astronomy, University of Michigan, 500 Church St., Ann Arbor, MI 48109, USA\\
$^{13}$Center for Astrophysical and Planetary Science, Racah Institute of Physics, The Hebrew University, Jerusalem 91904, Israel\\
$^{14}$Orange Coast College, Costa Mesa, CA 92626, USA\\
$^{15}$Space Telescope Science Institute, 3700 San Martin Drive, Baltimore, MD 21218, USA\\
$^{16}$The Observatories, The Carnegie Institution for Science, 813 Santa Barbara Street, Pasadena, CA 91101, USA\\
$^{17}$University of California, Riverside, 900 University Ave, Riverside, CA 92521, USA\\
$^{18}$Department of Physics and Astronomy, University of Pittsburgh, Pittsburgh, PA 15260, USA\\
$^{19}$Goddard Space Flight Center, Code 665, Greenbelt, MD, USA\\
$^{20}$Department of Physics and Astronomy, Texas A\&M University, College Station, TX, 77843-4242 USA\\
$^{21}$Max Planck Institute for Astronomy (MPIA), Konigstuhl 17, D-69117 Heidelberg, Germany\\

\end{document}